\def\C{{\mathchoice
{\setbox0=\hbox{$\displaystyle\rm C$}\hbox{\hbox to0pt
{\kern0.4\wd0\vrule height0.9\ht0\hss}\box0}}
{\setbox0=\hbox{$\textstyle\rm C$}\hbox{\hbox to0pt
{\kern0.4\wd0\vrule height0.9\ht0\hss}\box0}}
{\setbox0=\hbox{$\scriptstyle\rm C$}\hbox{\hbox to0pt
{\kern0.4\wd0\vrule height0.9\ht0\hss}\box0}}
{\setbox0=\hbox{$\scriptscriptstyle\rm C$}\hbox{\hbox to0pt
{\kern0.4\wd0\vrule height0.9\ht0\hss}\box0}}}}

\documentstyle[11pt]{article}
\begin{document}
\title{\mbox{ } \\[-3cm]
{\scriptsize\hspace*{\fill}May -- 1993 \\
 \scriptsize\hspace*{\fill}DF/IST 5.93 \\
 \scriptsize\hspace*{\fill}hep-th/9305173} \\[2cm]
  \LARGE   Chen Integrals, Generalized Loops and Loop Calculus}
\author{
{\Large J. N. Tavares}$^{1),2)}$
\\ \\
{\small 1) Dept.  Matem\'atica Pura, Fac. Ci\^encias, Univ. Porto, 4000
Porto,   PORTUGAL} \\
{\small 2) Dept.  F\'{\i}sica, Inst. Superior
T\'{e}cnico, 1096 Lisboa, PORTUGAL} \\
}
\date {\hspace{1cm}}
\maketitle
\begin{abstract}
  We use Chen iterated line integrals to construct a topological
algebra ${\cal A}_p$ of separating functions on the {\it Group of Loops} ${\bf
L}{\cal M}_p$. ${\cal A}_p$ has an Hopf algebra structure which allows the
construction of a group structure on its spectrum. We call this topological
group,  the group of generalized loops $\widetilde {{\bf L}{\cal M}_p}$.

Then we develope a {\it Loop Calculus}, based on the {\it
Endpoint} and {\it Area Derivative Operators}, providing a rigorous
mathematical treatment of early heuristic ideas of Gambini, Trias and also
Mandelstam, Makeenko and Migdal. Finally we define a natural action of the
"pointed" diffeomorphism group $Diff_p({\cal M})$ on $ \widetilde {{\bf L}{\cal
M}_p}$, and consider a {\it Variational Derivative} which allows the
construction of homotopy invariants.

This formalism is useful to construct a mathematical theory of {\it Loop
Representation} of Gauge Theories and Quantum Gravity.
\end{abstract}

\newpage
\pagestyle{plain}

\newtheorem{propo}{Proposition}

\section {Introduction}

\medskip
\medskip

   Let $\cal PM$ (resp. $\cal LM$) be the Path space (resp., the Loop space)
of a smooth manifold $\cal M$. On  $\cal PM$ (resp. $\cal LM$), we consider
a "sufficiently large" class of functionals, constructed through the so
called Chen iterated (line) integrals. This kind of integrals have been
introduced, a few years ago, by K.T.Chen, who used them extensively as an
useful tool in studying several algebraic topological aspects of Path
spaces (see [Chen 1,2,3,4]).

Here we consider them as a class of separating functions on the so called {\it
Group of Loops}  ${\bf L}{\cal M}_p$, consisting of equivalence classes of
picewise regular loops based at $p$, under a {\it "retracing"} equivalence.
This
group has been considered many years ago by Lefschetz and mainly by Teleman,
who has studied its representations to obtain a reconstruction theory for
principal fiber bundles with connection (see [T]). In this approach, ${\bf
L}{\cal M}_p$ plays the role of the fundamental group in classification of flat
bundles. More recently, Barrett and Lewandowsky have improved Teleman's
work establishing interesting connections with gauge theories (see [B],[Lew]).

We endow the algebra ${\cal A}_p$, generated by the above mentioned iterated
integrals, with the structure of topological algebra. This algebra has also a
natural Hopf algebra structure, which allows to introduce on its spectrum, a
structure of topological group. We call this group the {\it Group of
Generalized Loops} and we denote it by  $\widetilde {{\bf L}{\cal M}_p}$.
   It follows that  ${\bf L}{\cal M}_p$ is now embedded as a subgroup of
$\widetilde {{\bf L}{\cal M}_p}$, and so it is  a topological group, with the
induced topology. This topology is well fited to traduce the intuitive idea
that two loops are close if they differ by a "small area", and gives a correct
setting to develope several "directional" derivatives that are defined in
section 4. The relation of generalized loops to smooth loops resembles that of
distributions to smooth functions or that of DeRham currents to  smooth
chains.

The concept of generalized loop appears already in [Chen 5], where some
related algebraic aspects are considered.  Here, however, we use  affine group
theory [Ab], which substancially simplifies the treatment and allows us to
carry on  much farther the analogy with Lie Group Theory.

During the preparation of this work, we have received a preprint of Di
Bartolo, Gambini, Griego (see [BGG]), in which it is constructed an
{\it Extended Loop Group} that, apparently at least, has similarities with our
$\widetilde {{\bf L}{\cal M}_p}$. However (we believe that) these authors use a
substantially different formalism from ours, and thus, for the moment it is not
very clear for us, wether the two objects are really the same.

This work has grown as an effort to understand, from a mathematical viewpoint,
early work developed by Mandelstam, Makeenko, Migdal and mainly Gambini, Trias
(see [Man],[MM],[GT1],[GT2]). Thus, conceptually, most of the ideas here
formalized, are present in one or another form in the work of the above
mentioned authors. Here, however, the emphasis is in   a mathematical
framework which  can be used to construct a mathematical theory of loop
representation of gauge theories (see [L], for a recent review) and quantum
gravity (in Ashtekar-Rovelli-Smolin formulation, see [Ash],[RS],[G2]). In
particular, our constructions  are explicitly free of any background metric
structure and are really intrinsic.

 The present paper is organized as follows. In section 2, we
briefly review the definition and the main properties of Chen iterated line
integrals, establishing as well, their  relationship with geometry, via the
holonomy of a connection in a principal fiber bundle. In section 3, we define
a topological algebra ${\cal A}_p$, generated by the iterated integrals, which
furthermore has a natural Hopf algebra structure. Then, inspired by the theory
of affine groups, we define the group of generalized loops $\widetilde {{\bf
 L}{\cal
M}_p}$, as well as its Lie algebra $\widetilde {l{\cal {M}}_p}$. In section 4,
we develope a {\it Loop Calculus} based on the introduction of several
operators. So, we define {\it Endpoint Derivatives} and {\it Area Derivatives}
and establish sufficient machinery that turn them useful for efective
computations. At a conceptual level, most of these operators appear
heuristically formulated in early work of Mandelstam, Makeenko, Migdal and
mainly Gambini, Trias (we feel that our approach, in section 4, is more close
to that of these last two authors). However, here we put emphasis in the
mathematical foundations of what we call {\it Loop Calculus}. In fact, our aim
is to apply these loop calculus in a well formulated mathematical theory of
loop
representation of gauge theories and quantum gravity, subjects that are now
under investigation, and will be published elsewhere. Finally, in section 5, we
analyze the natural action of the "pointed" diffeomorphism group $Diff_p({\cal
M})$ on $\widetilde {{\bf L}{\cal
M}_p}$, and consider a {\it Variational Derivative} which allows the
construction of some simple homotopy invariants.
 We hope that this framework will be also useful in reconstruction theory of
principal fiber bundles with connection and  (why not?!) knot theory.

\medskip
\medskip
\medskip
\medskip

\section {Chen Iterated Line Integrals. Definition and main Properties}

\medskip
\medskip

   Let $\cal M$ be a smooth n-dimensional manifold. Denote by $\cal PM$ the
set of picewise smooth paths $\gamma:I\rightarrow {\cal M}$. Given real 1-forms
$\omega_1,...,\omega_r$ in $\bigwedge^1 {\cal M}$ and a path $\gamma \in
{\cal PM}$, we define the iterated (line) integrals inductivelly, as
follows:

\begin {eqnarray}
\int_{\gamma} \omega_1 &=& \int_{0}^{1} \omega_1 (t)dt {\nonumber}\\
\int_{\gamma} \omega_1 \omega_2 &=& \int_{0}^{1}
(\int_{0}^{t} \omega_1(s)ds) \omega_2 (t)dt {\nonumber}\\
&=&\int_{0}^{1}(\int_{\gamma^t} \omega_1) \omega_2 (t)dt {\nonumber}\\
&..........................& {\nonumber}\\
\int_{\gamma} \omega_1...\omega_r
&=&\int_{0}^{1} (\int_{\gamma^t} \omega_1...\omega_{r-1}) \omega_r (t)dt
\end {eqnarray}
where we have used the notations
$\omega_k(t) \equiv \omega_k(\gamma(t)).{\dot{\gamma}}(t)$ and
$\gamma^t:I\rightarrow
{\cal M}$ defined by $\gamma^t(s)\equiv \gamma(ts)$, for a fixed $t \in
I$.

Each iterated integral will be considered as a
function $X^{\omega_1...\omega_r}:{\cal PM} \rightarrow R$ defined by:

\begin {equation}
X^{\omega_1...\omega_r}(\gamma) = \int_{\gamma} \omega_1...\omega_r.
\end {equation}

Note that the above definitions work equally
well for 1-forms on $\cal M$ with values in an associative algebra $\bf A$
(p.ex., $C$ or  $gl(p)=gl(p,C)$, the algebra of $p \times p$
complex matrices). Of course in this case the functions
$X^{\omega_1...\omega_r}$ take values on $\bf A$. So, for example,
$X^{\omega_1\omega_2}(\alpha) =\int_{\alpha} \omega_1\omega_2$, with
$\omega_1,\omega_2 \in \bigwedge^1 {\cal M} \otimes gl(p)$ (i.e.,
$\omega_1,\omega_2$ are two matrices of 1-forms in $\cal M$), denotes the
matrix
in $gl(p)$:
\begin {eqnarray}
\left( \int_{\alpha} \omega_1\omega_2 \right)^i_j &=& \int_{\alpha}
(\omega_1)^i_k \otimes (\omega_2)^k_j \nonumber \\
                       &=& \int_{\alpha} (\omega_1)^i_k (\omega_2)^k_j
\end {eqnarray}
and analousgly for $\int \omega_1\omega_2...\omega_r$. (We have used notation
(5) in (3)).

\medskip

Let us now state the main properties of those iterated integrals.

\subsection {Proposition}
$\int_{\gamma} \omega_1...\omega_r$ is independent on orientation preserving
reparametrizations of $\gamma $.

\medskip
\medskip
\medskip

Before stating other properties, let us define the so called {\it Shuffle
Algebra of} $\cal M$. Let ${\bf k}=R$ or $C$. Consider the $\bf k$-vector
space $\bigwedge ^{1}{\cal M}$ of $\bf k$-1-forms on $\cal M$, and the tensor
algebra (over $\bf k$) of $\bigwedge ^{1}{\cal M}$:

\begin {equation}
{\cal T}({\bigwedge}^{1}{\cal M}) = \bigoplus _{r \geq 0}
(\bigotimes ^r {\bigwedge}^{1}{\cal M})
\end {equation}

For simplicity we use the notation:

\begin {equation}
\omega_1 ... \omega_r = \omega_1 \otimes...\otimes \omega_r \in
\bigotimes ^r {\bigwedge}^{1}{\cal M}
\end {equation}
 for $r \geq 1$, and set $\omega_1 ... \omega_r =1$, when $r=0$. Now we
replace the tensor multiplication in ${\cal T}(\bigwedge ^{1}{\cal M})$ by
the {\it shuffle multiplication} $\bullet$, defined by:

\begin {equation}
\omega_1 ... \omega_r \bullet \omega_{r+1} ... \omega_{r+s} =
{\sum}'_{\sigma} \omega_{\sigma (1)} ... \omega_{\sigma (r)}
\end {equation}
where $\sum '_{\sigma}$ denotes sum over all
$(r,s)$-shuffles, i.e., permutations $\sigma$ of $r+s$ letters with
$\sigma^{-1}(1) < ...< \sigma^{-1}(r)$ and $\sigma^{-1}(r+1) < ... <
\sigma^{-1}(r+s)$.
\medskip

{\small For example:

\begin {eqnarray*}
\omega_1 \bullet  \omega_2 &=& \omega_1 \omega_2 + \omega_2 \omega_1\\
\omega_1 \bullet  \omega_2 \omega_3 &=& \omega_1 \omega_2 \omega_3 +
\omega_2 \omega_1 \omega_3 + \omega_2 \omega_3 \omega_1.
\end {eqnarray*}}

(${\cal T}({\bigwedge}^{1}{\cal M}), \bullet$) is then an associative,
graded commutative $\bf k$-algebra, with unity $1 \in {\bf k} \subset {\cal
T}({\bigwedge}^{1}{\cal M})$, which is called the {\it Shuffle Algebra} of
$\cal M$ and is denoted by $Sh(\cal M)$, or simply by $Sh$.

\subsection {Proposition}

\begin {equation}
\int_{\gamma} \omega_1...\omega_r . \int_{\gamma}
\omega_{r+1}...\omega_{r+s} =   \int_{\gamma}
 \omega_1 ... \omega_r \bullet \omega_{r+1} ... \omega_{r+s}
\end {equation}
for $\omega_1,...,\omega_r \in \bigwedge
^{1}{\cal M}$.
\medskip

{\small For example, we have:

\begin {eqnarray*}
\int_{\gamma} \omega_1 . \int_{\gamma} \omega_2 &=& \int_{\gamma}\omega_1
\bullet \omega_2   \\
&=& \int_{\gamma} \omega_1 \omega_2 +
\omega_2\omega_1 \\
\int_{\gamma} \omega_1 . \int_{\gamma} \omega_2 \omega_3
&=&\int_{\gamma}\omega_1 \bullet \omega_2\omega_3 \\
&=& \int_{\gamma}\omega_1\omega_2 \omega_3 + \omega_2 \omega_1 \omega_3 +
\omega_2 \omega_3 \omega_1.
\end {eqnarray*}}

When $\omega_1,...,\omega_r \in \bigwedge
^{1}{\cal M} \otimes gl(p)$, the same formula (7) holds, but in the RHS of
(7), $\omega_1 ... \omega_r \bullet \omega_{r+1} ... \omega_{r+s}$ means the
product of the matrix $\omega_1 ... \omega_r$ by the matrix $\omega_1 ...
\omega_{r+s}$, the entries being multiplied through the shuffle product
$\bullet$ ($\omega_1 ... \omega_r$ means in turn, the product of the matrices
$\omega_1,\omega_2, ..., \omega_r$, the entries being now multiplied through
 $\otimes$).

\medskip
\medskip

\medskip
\medskip
\medskip
\medskip

{}From the equality
$f(\gamma(t))=f(\gamma(0))+\int_{{\gamma}^t}df$, we deduce the following:

\subsection {Proposition}

For any  $f \in C^{\infty}{\cal M}$ (resp. $f \in C^{\infty}{\cal M} \otimes
gl(p))$, we have:

\begin {eqnarray}
\int_{\gamma} df. \omega_1...\omega_r &=& \int_{\gamma}
(f.\omega_1) \omega_2...\omega_r - f(\gamma(0)). \int_{\gamma}
\omega_1...\omega_r \\
\int_{\gamma} \omega_1...\omega_r. df &=&
 \Big(\int_{\gamma} \omega_1...\omega_r \Big).f(\gamma(1))  - \int_{\gamma}
\omega_1...\omega_{r-1}.({\omega_r}.f) \nonumber\\
&& \\
\int_{\gamma} \omega_1...\omega_{i-1}. (df). \omega_{i+1}
..\omega_{r} &=&
\int_{\gamma}
\omega_1...\omega_{i-1}.(f.\omega_{i+1}).\omega_{i+2}...\omega_r
\nonumber \\
& & \ \ \ \  - \int_{\gamma}
\omega_1....(\omega_{i-1}.f).\omega_i...\omega_r \\
\int_{\gamma} \omega_1...\omega_{i-1}. (f. \omega_i). \omega_{i+1}
..\omega_{r} &=& f(\gamma(0)).  \int_{\gamma} \omega_1...\omega_r
\nonumber \\ && \ \ \ \  + \int_{\gamma} ((\omega_1 ...
\omega_{i-1})\bullet df). \omega_i ...\omega_{r}
\end {eqnarray}
for $\omega_1,...,\omega_r \in \bigwedge
^{1}{\cal M}$ (resp., $\in \bigwedge
^{1}{\cal M} \otimes gl(p)$).

\subsection {Proposition}
If $\alpha, \beta \in {\cal PM}$, with $\alpha (1)=\beta (0)$, then:

\begin {equation}
\int_{\alpha . \beta} \omega_1...\omega_r = \sum
_{i=0}^{r} \int_{\alpha} \omega_1...\omega_i .\int_{\beta}
\omega_{i+1}...\omega_r
\end {equation}
and,
\begin {equation}
\int_{\alpha ^{-1}}
\omega_1...\omega_r = (-1)^r \int_{\alpha} \omega_r...\omega_1
\end {equation}
for $\omega_1,...,\omega_r \in \bigwedge
^{1}{\cal M}$, and with the convention that $\int_{\gamma}
\omega_1...\omega_r = 1$, if $r=0$.

Moreover formula (12) holds with  $\omega_1,...,\omega_r \in \bigwedge
^{1}{\cal M} \otimes gl(p)$, while formula (13) must be substituted by:
\begin {equation}
\int_{\alpha ^{-1}}
\omega_1...\omega_r = (-1)^r \int_{\alpha}
[\omega_r^{T}...\omega_1^{T}]^{T}
\end {equation}
where $\omega^{T}$ means transpose of the matrix $\omega$  (note that here
$A^{T}B^{T}$ is not equal to $[BA]^{T}$).

\medskip
\medskip

Now we analyse the separating properties of the functions
$X^{\omega_1...\omega_r}:{\cal PM} \rightarrow R$, defined by (2). We shall
not distinguish two paths in ${\cal PM}$, which differ only by orientation
preserving reparametrization.

Two paths are called {\it {elementary equivalent}} if one of them can be
written in the form $\alpha \beta \beta ^{-1} \gamma$, and the other in the
form $\alpha \gamma$. If $\alpha _1,...,\alpha_s$ is a finite sequence of
paths, such that $\alpha_i$ and $\alpha _{i+1}$ are elementary equivalent
($i=1,...,s-1$), then we say that $\alpha_1$ is {\it{equivalent}} to
$\alpha_s$. We denote by $[\alpha]$ the equivalence class of $\alpha$. A
{\it {picewise regular path }}is a path in $\cal PM$ with nonvanishing
tangent vectors. Finally, a {\it {reduced path}} is a picewise regular path
that is not of the type $\alpha \beta \beta ^{-1} \gamma$, for any $\beta$.

It's easy to see that the functions $X^{\omega_1...\omega_r}$ depend only
on the equivalence class $[\alpha]$ of the path $\alpha$. The following
lemma is proved in [Chen 3]:

\subsection {Lemma}

Let $\alpha$ be a nonempty reduced picewise regular path in ${\cal PM}$.
Then there exists 1-forms $\omega_1,...,\omega_r$ in $\bigwedge
^{1}{\cal M}$, $r \geq 1$, such that:

\begin {equation}
X^{\omega_1...\omega_r} (\alpha) \neq 0 \nonumber
\end {equation}

As a consequence, it was also proved that every picewise regular path is
equivalent to one and only one reduced path. The next theorem answers the
question how well iterated integrals separate paths (see [Chen 4]):

\subsection {Theorem}

Two picewise regular paths $\alpha$, $\beta$ are equivalent if and only if

\begin {equation}
X^{\omega_1...\omega_r} (\alpha) = X^{\omega_1...\omega_r} (\beta)
\nonumber  \end {equation}
for any 1-forms
$\omega_1,...,\omega_r$ in  $\bigwedge ^{1}{\cal M}$, \ \  $r \geq 1$.

\medskip

\underline {Proof}...

{\small $(\Rightarrow)$...definitions.

$(\Leftarrow)$... By the above lemma, we may assume that $X^{\omega_1...
\omega_r} (\alpha) \neq
0$, for some $\omega_1,...,\omega_r$ in  $\bigwedge ^{1}{\cal M}$, $r
\geq 1$. Otherwise, both $\alpha$ and $\beta$ are equivalent to the
empty reduced path, and the theorem follows.

Given any $f \in C^{\infty}{\cal M}$, we obtain from (8):

\begin {equation}
\int_{\alpha} (f\omega_1) \omega_2...\omega_r =
\int_{\alpha} df \omega_1...\omega_r +  f(\alpha(0))
\int_{\alpha} \omega_1...\omega_r
\end {equation}
and an analogous formula for $\beta$ holds, by the hypothesis. It follows
that $f(\alpha(0)) = f(\beta(0))$, $\forall f \in C^{\infty}{\cal M}$, and
so $\alpha(0) = \beta (0)$.

Now, $\beta^{-1} \alpha$ is a picewise regular path and using (12-13),
we verify that every iterated integral vanishes along $\beta^{-1}
\alpha$, which implies, again by the lemma, that $\beta^{-1} \alpha$
is equivalent to the reduced empty path, and so $\alpha \sim \beta$,
QED.}

\medskip
\medskip

There is an interesting relation between iterated integrals and geometry,
via the parallel transport of a connection on a trivial bundle, that we
can use to prove the above propositions. In fact, assume that $\nabla$ is
a connection on the trivial bundle $R^p \times {\cal M} \rightarrow
{\cal M}$, over ${\cal M}$. Sections of this bundle are identified with
functions $s :{\cal M} \rightarrow R^p$, and its canonical framing $e$ is
given by the $p$ constant functions:

\begin {equation}
e_i :M \rightarrow R^p
\end {equation}
$i=1,...,p$, which take each point of ${\cal M}$ to the $i^{th}$ standard
basis vector of $R^p$.
So, for a section $s :{\cal M} \rightarrow R^p$, we can write $s= \sum s^i
e_i$, with $s^i \in C^{\infty} {\cal M}$.

As $\nabla e_i$ must be an $R^p$ -valued 1-form on $\cal M$, we can write:
\begin {equation}
\nabla e_i = - \sum \omega _i^j e_j
\end {equation}
where $\omega =(\omega _i^j)$, a $p \times p$ matrix of 1-forms on $\cal
M$, is the connection form (associated to the framing $e$ and the
connection $\nabla$). Now, for a section $s$ as above, we have:

\begin {eqnarray}
\nabla s &=& \nabla (s^i e_i) = s^i \nabla e_i + ds^i. e_i  \nonumber \\
         &=& s^i (-\omega _i^j e_j) + ds^i. e_i \nonumber \\
         &=& (ds^j - s^i \omega _i^j)e_j \nonumber \\
         &=& (ds - s. \omega)^j e_j
\end {eqnarray}
i.e., in matrix notation ($s$ as a line-vector):
\begin {equation}
\nabla s = ds - s. \omega.
\end {equation}

Conversely, if $\omega \in \bigwedge ^{1}{\cal M} \otimes gl(p)$
is a $p \times p$ matrix of 1-forms, we can define a connection on $R^p
\times {\cal M} \rightarrow {\cal M}$, by (21).

This connection lifts to the principal fiber bundle $Gl(p) \times {\cal M}
\rightarrow {\cal M}$, by defining:

\begin {equation}
\nabla S = dS - S. \omega
\end{equation}
where $S:{\cal M} \rightarrow Gl(p)$. A section $S: I \rightarrow Gl(p)$,
along a path $\gamma :I \rightarrow {\cal M}$, is called {\it {horizontal}}
if:

\begin {equation}
dS(t) = S(t). \gamma^* \omega
\end {equation}
$\forall t \in I$. Denoting by $ A(t)dt = \gamma^* \omega $, (23)
becomes: \begin {equation}
S'(t) = S(t).A(t).
\end{equation}

We define the parallel transport as the map $U: {\cal PM} \rightarrow
Gl(p)$, $\gamma \mapsto U_{\gamma}$, defined as follows: $U_{\gamma} =
S(1)$, where $S: I=[0,1] \rightarrow Gl(p)$ is the unique horizontal
section along $\gamma$, i.e., the unique solution of the linear
(nonautonomous) differential equation (24), satisfying $S(0)=Id$. By
the general theory of linear ordinary differential equations, we know that
$U_{\gamma}$ is independent of the parametrization of $\gamma$ and, if $\alpha
, \beta \in {\cal PM}$, with $\alpha (1) = \beta (0)$, then $U_{\alpha .\beta}
=
U_{\alpha} U_{\beta}$.

We can give a formula for $U_{\gamma}$ in terms of iterated integrals
of the connection form $\omega = (\omega ^i_j)$. First it's easy to prove
that, for $\gamma \in {\cal PM}$, there exists a constant $M>0$ such that:

\begin {equation}
\parallel \int_{\gamma} \underbrace{ \omega \omega \cdots \omega}_{r}
 \parallel = O( \frac {M^r}{r!})
\end {equation}
and so, the series:
\begin {equation}
Id + \int_{\gamma} \omega + \int_{\gamma} \omega \omega + \int_{\gamma} \omega
\omega \omega
+ ...
\end {equation}
converges in $Gl(p)$. Moreover, $U_{\gamma}$ is given exactly by this
"chronological series" of iterated integrals (see [DF] for all this):
\begin {equation}
U_{\gamma} =
Id + \int_{\gamma} \omega + \int_{\gamma} \omega \omega +
\int_{\gamma} \omega \omega \omega
+ ...
\end {equation}

Of special importance is the case where $\omega  \in \bigwedge ^{1}{\cal
M} \otimes {\cal N}_r$, where ${\cal N}_r$ denotes the Lie algebra of
nilpotent upper triangular $(r+1) \times (r+1)$ matrices. In this case,
the sum in (27) is finite. For example, if $\omega_1,...,\omega_r \in
\bigwedge ^{1}{\cal M}$ and:

$$\omega = \left[
\begin{array}{cccccc}
0  &  \omega_1  & 0        &  \ldots & \ldots & 0        \\
 0  &     0      & \omega_2 & 0       & \ldots & 0        \\
 \ldots  & \ldots & \ldots  & \ldots & \ldots  & \ldots   \\
0  &     0      &    0     &   \ldots      &    0    &  \omega_r \\
0  &    0      &    0     &   \ldots      &    0    & 0
\end{array}
\right]$$
then,
$$U = \left[
\begin{array}{cccccc}
1   &  \int \omega_1  & \int \omega_1 \omega_2  & \int \omega_1 \omega_2
\omega_3 & \ldots & \int \omega_1 \omega_2 ... \omega_r \\
 0  &     1      & \int \omega_2 & \int \omega_2 \omega_3  & \ldots & \int
\omega_2...\omega_r \\
  0 &       0      & 1   &  \int \omega_3 &  \ldots   & \int \omega_3 ...
\omega_r  \\
\ldots & \ldots &\ldots &\ldots &\ldots &\ldots \\
 0 &     0       &    0     & \ldots  & 1  & \int \omega_r  \\
0 & 0 & 0 & \ldots  & 0  & 1
\end{array}
\right]$$

We can use this, to prove most of the above mentioned properties of
iterated integrals.

If $P({\cal M},G)$ is a  principal fiber bundle over
$\cal M$, with structure group $G$ (a subgroup of $Gl(p)$), we can apply
the same reasoning in each trivializing chart of that bundle. If $\gamma$
is a loop in $\cal M$, we obtain, by the above construction, a (gauge
dependent) transformation $U_{\gamma} \in G$, which is called the {\it
Holonomy} of the connection $\omega$ around the loop $\gamma$.

Under a gauge transformation $g:{\cal U} \subseteq {\cal M} \rightarrow G$,
we have that: \begin {equation}
U_{\gamma}^g = g_x^{-1} U_{\gamma} g_x
\end {equation}
where $x=\gamma(o)$, and so, we obtain a gauge independent loop functional
${\cal W}:{\cal LM} \rightarrow C$, defined by:
\begin {equation}
{\cal W}(\gamma) = Trace \, U_{\gamma}
\end {equation}
which is usually called {\it Wilson loop variable}.

By continuity of the trace, we obtain from (27):
\begin {equation}
{\cal W}(\gamma) = \sum_{r \geq 0} Trace \int_{\gamma} \underbrace{
\omega \omega \cdots \omega}_{r}
\end {equation}
with the convention $\int_{\gamma} \underbrace{ \omega \omega \cdots
\omega}_{r}=Id$, if $r=0$.
So ${\cal W}(\gamma)$ is given by a convergent series of cyclic
combinations of iterated integrals of the 1-forms $\omega^i_j$ in the
connection matrix $\omega = (\omega^i_j)$.

\section {The Group of Generalized Loops and its Lie Algebra}

\subsection {The Shuffle Algebra over $\bigwedge ^{1}{\cal M}$, as an Hopf
Algebra }

 Recall that we have defined the {\it Shuffle Algebra} of
$\cal M$,  denoted by $Sh(\cal M)$, or simply by $Sh$, as
(${\cal T}({\bigwedge}^{1}{\cal M}), \bullet$). $Sh$ is then an associative,
graded, commutative $\bf k$-algebra, with unity $1 \in {\bf k} \subset {\cal
T}({\bigwedge}^{1}{\cal M})$.

$Sh$ has also a $\bf k$-Hopf algebra structure. This means
(see [Ab], [Sw, Chp.XII]) that, in adition to the above $\bf k$-algebra
stucture, we have two $\bf k$-linear maps $\Delta :Sh \rightarrow Sh \otimes
Sh$, called {\it comultiplication}, and $\epsilon :Sh \rightarrow {\bf k}$,
called {\it counity}, defined respectivelly, by the formulas:
\begin {eqnarray}
\Delta (\omega_1 ... \omega_r) &=& \sum_{i=0}^{r} \omega_1 ...
\omega_i \otimes \omega_{i+1} ... \omega_r \\
\epsilon (\omega_1 ... \omega_r) &=& 0, \ \ \ if \ \ r \geq 1 \\
 &=& 1,  \ \ \ if \ \ r=0
\end {eqnarray}
which verifies the following identities:
\begin {eqnarray}
(\Delta \otimes 1) \circ \Delta &=& (1 \otimes \Delta) \circ \Delta \ \
\ \ \ ({\hbox {Coassociative law}}) \\
(1 \otimes \epsilon) \circ \Delta &=& (\epsilon \otimes 1) \circ \Delta = 1
 \ \ \ \ ({\hbox {counitary property}}) \\
\Delta({\bf u} \bullet {\bf v}) &=& \Delta({\bf u}) \bullet \Delta({\bf v}) \ \
\ \ \ ({\hbox {$\Delta$ is an algebra morphism}}) \\
\epsilon({\bf u} \bullet {\bf v}) &=& \epsilon({\bf u}) \bullet
\epsilon({\bf v}) \ \ \ \ \ ({\hbox {$\epsilon$ is an algebra morphism}})
\end {eqnarray}
$\forall {\bf u},{\bf v} \in Sh$.

Moreover there is also a $\bf k$-linear map $J: Sh \rightarrow Sh$, called
{\it antipode}, defined by:
\begin {equation}
J(\omega_1 ... \omega_r)=(-1)^r \omega_r ... \omega_1
\end {equation}
which verifies:
\begin {eqnarray}
s \circ (J \otimes 1) \circ \Delta &=& s \circ (1
\otimes J) \circ \Delta = \eta \circ \epsilon \\
J({\bf u} \bullet {\bf v}) &=& J({\bf v}) \bullet J({\bf u}) \\
J(1) &=& 1 \\
\epsilon \circ J &=& \epsilon \\
\tau \circ (J \otimes J) \circ \Delta &=& \Delta \circ J \\
J^2 &=& 1
\end {eqnarray}
$\forall {\bf u},{\bf v} \in Sh$. In (39),   $s :Sh \otimes Sh
\rightarrow Sh$ denotes shuffle multiplication,
and    $\eta :{\bf k} \rightarrow Sh$ the unit map.
Finally, in (43), $\tau: Sh \otimes Sh \rightarrow
Sh \otimes Sh $ is the transposition map $\tau ({\bf u}
\otimes {\bf v})= {\bf v} \otimes {\bf u}$.

 Note that (39) means explicitly the following identity:

\begin {eqnarray}
\sum_{i=0}^{r}(-1)^i \omega_i ... \omega_1 \bullet \omega_{i+1} ...
\omega_r &=& \sum_{i=0}^{r}(-1)^{r-i} \omega_1 ... \omega_i \bullet
\omega_{r} ... \omega_{i+1} \nonumber \\
&=& \epsilon (\omega_1 ... \omega_r)
\end {eqnarray}

\medskip
\medskip

We endow $Sh(\cal M)$ with the structure of nuclear LMC
algebra (see [Mal]) in the following way.

First we topologize the vector space ${\bigwedge}^{1}{\cal M}$. Given a
local chart  $({\cal U},x)$ of $\cal M$, choose a nested sequence of compacts
$\{K_m^{\cal U}\}_{m \geq 1}$, in $\cal U$, such that ${\cup}_{m \geq 1}
 K_m^{\cal U}
= {\cal U}$ (this is always possible). Then  we topologize the vector space
${\bigwedge}^{1}{\cal U}$ through the familly of seminorms:
\begin {equation}
N^{\cal U}_{m}(\omega) =  {\max_{1 \leq i \leq n}} \rho^{\cal
U}_{m}(\omega_i) \ \ \ \  \omega \in {\bigwedge}^{1}{\cal U}
\end {equation}
where $m$ is a positive integer, $\omega = \sum_{i=1}^{n} \omega_{i}dx^i$  and:
\begin {equation}
\rho^{\cal U}_{m}(\omega_i)= \sup_{|p| \leq m}(\sup_{x \in K_m^{\cal U}}|D^p
{\omega_i}(x)|)
\end {equation}
for ${\omega_i} \in C^{\infty}{\cal U}$.

Each ${\bigwedge}^{1}{\cal U}$ becomes in this way a nuclear locally
convex topological vector space.

Now, let $\{{\cal U}_k\}_{k \in N}$ be a basis for the topology of $\cal M$,
consisting of local charts (this is always possible). The inclusions
$i_k:{\cal U}_k \rightarrow {\cal M}$, induce linear maps
$i_k^{*}:{\bigwedge}^{1}{\cal M} \rightarrow {\bigwedge}^{1}{\cal U}_k$. We
endow ${\bigwedge}^{1}{\cal M}$ with the initial topology defined by this maps.
By definition this is the weakest topology for which all the maps $i_k^{*}$ are
continuous, and a local basis consists of sets of the form $\cap _{j=1}^r
(i^{*}_{k_j})^{-1}({\cal O}_{k_j})$, where the sets ${\cal O}_{k_j}$ run over
local basis of ${\bigwedge}^{1}{\cal U}_{k_j}$. So,  ${\bigwedge}^{1}{\cal M}$
becomes in this way a nuclear locally convex topological vector space, whose
topology can be described by the family of seminorms:
\begin {equation}
p_{k,m,l}(\omega) = \max _{1 \leq j \leq l} N_m^{{\cal U}_{k_j}}\big(i_{{\cal
U}_{k_j}}^{*}\omega \big)
\end {equation}

In this topology, a sequence of 1-forms $(\omega_k)_{k \geq 1}$, in
 ${\bigwedge}^{1}{\cal
M}$, converges to zero iff, in a neighbourhood of each point of $\cal M$, each
derivative of each coeficient of $\omega_k$ converges uniformly to zero.

  Then, each tensor power $\bigotimes ^r {\bigwedge}^{1}{\cal M}$ is
topologized
through the so called projective tensor product topology, i.e., through the
seminorms $N^{(r)}_{k,m,l}$ which are the tensor product of the above ones (see
[Mal], chpt X). For example, for an ${\bf u} \in  {\bigwedge}^{1}{\cal M}
\otimes {\bigwedge}^{1}{\cal M}$ we have:
\begin {equation}
N^{(2)}_{k,m,l}({\bf u})=inf \sum_{i=1}^{n}
p_{k,m,l}(\omega^i).p_{k,m,l}(\eta^i)
\end {equation}
where $inf$ is taken over all expressions of the element $\bf u$ in the form
${\bf u}=\sum_{i=1}^{n} \omega^i \otimes \eta^i$. Finally, since an element in
$\bigoplus _{r \geq 0} (\bigotimes ^r {\bigwedge}^{1}{\cal M})$ is a
finite sum ${\bf u}=\sum_{r} {\bf u}_r$, with ${\bf u}_r \in \bigotimes ^r
{\bigwedge}^{1}{\cal M}$, we use the seminorms:
\begin {equation}
{\bf N}_{k,m,l}({\bf u})=\sum_{r} N^{(r)}_{k,m,l}({\bf u}_r)
\end {equation}
to put a locally convex topology in ${\cal T}({\bigwedge}^{1}{\cal M})$.
Since the shuffle product is continuous, we obtain in this way a
commutative LMC algebra which we continue to denote by $Sh(\cal M)$, or
simply by $Sh$.

\subsection {The group ${{\bf L}{\cal M}_p}$ of Loops and the Algebra of
Iterated Integrals}

Fix a point $p \in {\cal M}$, and consider the Loop Space ${\cal LM}_p$ of
picewise smooth loops based at $p$.  ${\cal LM}_p$ is a semigroup with respect
to the operation of justaposition of loops: $\alpha . \beta$, with $\alpha,
\beta \in {\cal LM}_p$. Recall that $[\alpha]$ denotes the equivalence class of
the loop $\alpha$, under the equivalence relation $\sim$, defined in
section 2.

Define an operation on the set ${\cal
LM}_p/{\sim}$, by:
\begin {equation}
[\alpha] \diamond [\beta] = [\alpha . \beta]
\end {equation}
which endows ${\cal LM}_p/{\sim}$ with the structure of group. The inverse
of an element $[\alpha] \in {\cal LM}_p/{\sim}$, is given by
$[\alpha]^{-1} = [\alpha ^{-1}]$, and the unit element is given by
$[{p}]$, the class of the constant loop equal to the point $p$. This
group (${\cal LM}_p/{\sim}, \diamond$) is called the {\it Group of loops}
of the manifold $\cal M$, based at $p$, and is denoted by ${\bf L}{\cal
M}_p$.
\medskip
\medskip

 Let ${\cal A}_p$ denote the algebra generated by
all the functions $X^{\omega_1...\omega_r}$, defined by (2) and considered
as functions on ${\bf L}{\cal M}_p$ . Then proposition 2.2 says that the
surjective map $Sh({\cal M}) \rightarrow {\cal A}_p$, defined by $1 \mapsto
1$ and $\omega_1...\omega_r \mapsto X^{\omega_1...\omega_r}$ is a morphism
of algebras. Moreover, Proposition 2.3 implies that the Kernel of this
morphism, contains the ideal ${\bf I}_p$ generated by all the elements of
type:
\begin {equation}
\omega_1...\omega_{i-1} (f \omega_i) \omega_{i+1}...\omega_{r}
- f(p)  \omega_1...\omega_r -  ((\omega_1
.. \omega_{i-1})\bullet df) \omega_i ...\omega_{r}.
\end {equation}
or, in short, by the elements of type:
\begin {equation}
{\bf u}(f\omega){\bf v} - ({\bf u} \bullet df)\omega {\bf v} - f(p){\bf
u}\omega {\bf v} \end {equation}
for ${\bf u},{\bf v} \in Sh, \ \ \omega \in {\bigwedge}^{1}{\cal M}, \ \ f \in
C^{\infty}{\cal M}$.

On the other hand, since $\int_{\gamma} df=0$, for a $\gamma \in {\cal
LM}$, this kernel also contains  $dC^{\infty}({\cal M})$. Denoting by
$<dC>$ the ideal generated by $dC^{\infty}({\cal M})$, in $Sh({\cal M})$, and
putting: \begin {equation}
{\bf J}_p = {\bf I}_p + <dC>
\end {equation}
then we have an algebra isomorphism (see [Chen 4]):
\begin {equation}
Sh({\cal M})/ {\bf J}_p \simeq {\cal A}_p.
\end {equation}

\medskip
\medskip

The algebra ${\cal A}_p$ admits  also a {\bf k}-Hopf Algebra structure, by
defining the comultiplication $\Delta : {\cal A}_p \rightarrow {\cal A}_p
\otimes {\cal A}_p$, the counity $\epsilon : {\cal A}_p \rightarrow {\bf
k}$ and the antipode $J:{\cal A}_p \rightarrow {\cal A}_p$ , respectivelly by:
\begin {eqnarray}
\Delta \big( X^{\omega_1...\omega_r}\big) &=& \sum_{i=0}^{r}
X^{\omega_1...\omega_i} \otimes X^{\omega_{i+1}...\omega_r}\\
\epsilon \big( X^{\omega_1...\omega_r}\big) &=& 0 \ \ \ \ \ if \ \ r \geq 1
\\ &=& 1 \ \ \ \ \ if \ \ r=0 \\
J \big( X^{\omega_1...\omega_r}\big) &=& (-1)^r
X^{\omega_r...\omega_1}
\end {eqnarray}
which verify the above formal properties (34) to (37), and (39) to (44).

\medskip
\medskip

Finally, we topologize the algebra ${\cal A}_p$ through the isomorphism (55),
obtaining, in this way, a commutative LMC algebra, generated by the functions
$X^{\omega_1...\omega_r}:{\bf L}{\cal M}_p \rightarrow {\bf k}$, which,
moreover, has an aditional (commutative, noncocomutative) Hopf algebra
structure. This is all we need to define the group of generalized loops, in the
next section.

\subsection {The Group ${\widetilde {{\bf L}{\cal M}_p}}$ of Generalized
Loops}

 Let us consider the spectrum ${\bf {\Delta}}_p$ (or Gelfand space) of the
algebra ${\cal A}_p$. By definition, ${\bf {\Delta}}_p$ consists of all
nonnull continuous algebra homomorphisms (characters) $\phi:{\cal A}_p
\rightarrow C$, endowed with the induced weak $\star$-topology (or Gelfand
topology). Equivalently, a character of  ${\cal A}_p$, is a continuous
complex algebra homomorphism $\tilde {\alpha} :Sh({\cal M}) \rightarrow C$,
that vanishes on the closed ideal ${\bf J}_p$, given by (54).

\subsubsection {Definition}
A {\it Generalized Loop} based at $p \in {\cal M}$ is a character of the
algebra ${\cal A}_p$ or, equivalently, a continuous complex algebra
homomorphism $\tilde {\alpha} :Sh({\cal M}) \rightarrow C$, that vanishes on
the ideal ${\bf J}_p$.

\medskip
\medskip

So, the set of all generalized loops is ${\bf \Delta}_p$, with the Gelfand
topology: a sequence $(\tilde {\alpha}_n)$ converges to $\tilde {\alpha}$,
in ${\bf {\Delta}}_p$, iff:
\begin {equation}
\lim_{n \rightarrow \infty} \tilde {\alpha}_n
(X^{\bf u}) = \tilde {\alpha}(X^{\bf u}) \ \ \ \ \ \ \forall {\bf u} \in
Sh({\cal M}).
\end {equation}

\medskip
\medskip

We have a natural embedding of ${\bf L}{\cal M}_p$ into  ${\bf
{\Delta}}_p$, given by the "Dirac map" $\delta:{\bf L}{\cal M}_p
\rightarrow {\bf {\Delta}}_p$, $[\alpha] \mapsto \delta_{[\alpha]}$ defined
by:  \begin {equation}
\delta_{[\alpha]}(X^{\omega_1...\omega_r}) =
X^{\omega_1...\omega_r}([\alpha]) \ \ \ \ [\alpha] \in {\bf L}{\cal M}_p.
\end {equation}

Since the functions $X^{\omega_1...\omega_r}$ separate "points" in ${\bf
L}{\cal M}_p$ (Theorem 2.6), we see that this is an injective embedding.
So, we identify ${\bf L}{\cal M}_p$ with its image, under $\delta$, in
${\bf {\Delta}}_p$, and endow ${\bf L}{\cal M}_p$ with the induced topology.
In this topology, a sequence $([\alpha]_n)$ converges to $[\alpha]$, in
${\bf L}{\cal M}_p$  iff $\lim_{n\rightarrow \infty}X^{\bf u}([\alpha]_n) =
X^{\bf u}([\alpha]), \ \ \  \forall {\bf u} \in Sh{\cal M}$.

\medskip
\medskip

Now we want to define a group structure in the set  ${\bf {\Delta}}_p$ of
generalized loops, and, for this, we use some facts about
$\bf k$-affine groups that can be seen in detail in [Ab].

 Thus, we define the group product ${\tilde {\alpha}} \star {\tilde
{\beta}}$ through the so called {\it convolution} of the two elements
${\tilde {\alpha}}, {\tilde {\beta}} \in {\bf {\Delta}}_p$. Let us explain
what this means.  Denoting by
 ${\cal A}_p^{*}$ the {\bf k}-dual linear space of ${\cal A}_p$,
we define a multiplication on ${\cal A}_p^{*}$ in the following way.
First note that ${\cal A}_p^{*} \otimes {\cal A}_p^{*} \cong \big( {\cal
A}_p \otimes {\cal A}_p \big) ^{*}$ and then, compose with the dual of the
comultiplication map,  to obtain the required
multiplication as follows:
\begin {equation}
{\cal A}_p^{*} \otimes {\cal A}_p^{*} \cong \big( {\cal
A}_p \otimes {\cal A}_p \big) ^{*} \stackrel {\Delta^{*}} \longrightarrow
{\cal A}_p^{*} \end {equation}

This multiplication is what we call the {\it convolution}, and is given by:

\begin {equation}
{\tilde {\alpha}} \star {\tilde {\beta}} = ({\tilde {\alpha}} \otimes
{\tilde {\beta}}) \circ \Delta
\end {equation}
where we have used the identification ${\bf k}\otimes {\bf k} \simeq {\bf
k}$ . More explicitly:
 \begin {equation}
{\tilde {\alpha}} \star {\tilde {\beta}}(X^{\omega_1...\omega_r}) = \sum
_{i=0}^{r} {\tilde {\alpha}} (X^{\omega_1...\omega_i}) .{\tilde
{\beta}}(X^{\omega_{i+1}...\omega_r}).
\end {equation}

We define also the inverse of ${\tilde {\alpha}} \in {\bf {\Delta}}_p$, by
${\tilde {\alpha}} \circ J$, i.e.:
 \begin {equation}
{\tilde {\alpha}}^{-1}(\omega_1...\omega_r) = (-1)^r {\tilde
{\alpha}}(\omega_r...\omega_1)
\end {equation}
and  take  $\epsilon$, given by (57-58), as the unit element

Now, the algebraic part of the following theorem follows directly from [Ab,
Th.2.1.5], while the topological part is of easy verification.

\subsubsection {Theorem}
${\tilde {\alpha}} \star {\tilde {\beta}}$, ${\tilde {\alpha}}^{-1}$ and
$\epsilon$ are generalized loops based at $p$, i.e., they are continuous
characteres on the algebra ${\cal A}_p$, or equivalently, continuous
characteres on the algebra
$Sh({\cal M})$ that vanish on the ideal ${\bf J}_p$.

Moreover, $({\bf {\Delta}}_p,.)$ has the structure of topological (Hausdorff
and completely regular) group.

\medskip
\medskip

\subsubsection {Definition}
We call the above mentioned topological group  $({\bf {\Delta}}_p,.)$, the
{\it Group of Generalized Loops} of $\cal M$, based at $p \in {\cal M}$,
and we denote it by ${\widetilde {{\bf L}{\cal M}_p}}$.

\medskip
\medskip
Note that under the identification given by the "Dirac map" $\delta:{\bf
L}{\cal
 M}_p
\rightarrow {\bf {\Delta}}_p$, (see  (61)), ${\bf L}{\cal M}_p$ is a
topological subgroup of ${\widetilde {{\bf L}{\cal M}_p}}$.

\subsubsection {Note}

{\small Each $X^{\bf u} \in {\cal A}_p$ defines a continuous function
$F_{X^{\bf u}}:{\widetilde {{\bf L}{\cal M}_p}} \rightarrow {\bf k}$ through:
\begin {eqnarray*}
F_{X^{\bf u}}(\tilde{\alpha}) \equiv X^{\bf u}(\tilde{\alpha})
\end {eqnarray*}

Each such function is a {\it representative function} of ${\widetilde {{\bf
 L}{\cal
M}_p}}$, i.e., the linear space generated by all left translates ${\tilde
{\alpha}}.F_{X^{\bf u}}$ is finite dimensional. In fact:
\begin {eqnarray*}
{\tilde{\alpha}}.F_{X^{\bf u}}=F_{(1 \otimes {\tilde {\alpha}})\Delta {X^{\bf
u}}}
\end {eqnarray*}
and, for each ${X^{\bf u}}$ fixed, the linear space generated by:
\begin {eqnarray*}
\{(1 \otimes {\tilde {\alpha}})\Delta {X^{\bf
u}}: {\tilde {\alpha}} \in {\widetilde {{\bf L}{\cal
M}_p}}\}
\end {eqnarray*}
is finite dimensional.}

\subsubsection {Example}

{\small Let ${\cal M}$=$S^1$. Then it's easy to see that ${\bf L}S^1_p=Z$.
However
 $\widetilde {{\bf L}{\cal S}}_p^1=R$. In fact, since $H^1(S^1,R)=R$, any
1-form $\omega$ in $S^1$ is equal to a constant multiple of ${\omega}_0 \equiv
d\theta$ (the usual volume form in $S^1$), modulo an exact form: $\omega =
c{\omega}_0 + df, c \in R$. So:
\begin {equation}
{\wedge}^{1}S^{1}=R{\omega}_0 \oplus dC^{\infty}(S^{1})
\end {equation}

{}From this fact, and using the relations (5-8), we can prove that ${\cal A}_p$
is isomorphic, as an Hopf algebra, to $R[t]$, the polynomial ring in one
variable $t \leftrightarrow X^{{\omega}_0}$ (see [Chen5]). The Hopf operations
on $R[t]$ are: \begin {equation}
\Delta(t)=1\otimes t + t \otimes1 \ \ \ J(t)=-t \ \ \ \epsilon(t)=0
\end {equation}

Now,  $\widetilde {{\bf L}{\cal S}}_p^1=R$, follows from Example 4.1 in
[Ab, pag.172].}

\subsubsection  {Note}

{\small Consider the Path space ${\cal PM}_p$ of paths based at $p \in {\cal
M}$, and the algebra ${\cal B}_p$ generated by all the functions
$X^{\omega_1...\omega_r}$, considered now as functions on ${\cal PM}_p$.
In exactly the same way as in the previous case, (see [Chen 4]) there is an
algebra isomorphism $Sh({\cal M})/{\bf I}_p \simeq {\cal B}_p$, which allows
to consider ${\cal B}_p$ as an LMC algebra and define {\it generalized
paths}, based at $p$, as continuous characteres on $Sh({\cal M})$, that
vanish on ${\bf I}_p$.}

\medskip
\medskip

\subsection { The Lie Algebra of the Group ${\widetilde {{\bf L}{\cal
M}_p}}$}

Consider the topological algebra ${\cal A}_p$ of iterated integrals with the
structure of commutative Hopf algebra described above.
 A {\bf k}-linear map $D:{\cal A}_p \rightarrow {\cal A}_p$ is called a {\it
Left Invariant Derivation} (resp., {\it Right Invariant Derivation}) on ${\cal
A}_p$ if $D$ verifies the following two conditions:
\begin {eqnarray}
D\big( X^{\bf u}X^{\bf v}\big) &=& X^{\bf u} D\big( X^{\bf v} \big) + D\big(
X^{\bf u} \big)X^{\bf v} \\ \Delta \circ D &=& (1 \otimes D) \circ \Delta.
\end {eqnarray}
(resp., $\Delta \circ D = (  D  \otimes 1) \circ \Delta$),
for all ${\bf u},{\bf v} \in Sh$.

The next definition follows from the general theory of affine $\bf
k$-groups (see [Ab, Chp.4.3]):

\subsubsection {Definition}

 We define the Lie Algebra of the Group ${\widetilde {{\bf L}{\cal
M}_p}}$ as the {\bf k}-linear space $\widetilde {l{\cal {M}}_p}$ of all
continuous  Left Invariant Derivations on ${\cal A}_p$.

\medskip

Of course the brackett in  $\widetilde {l{\cal {M}}_p}$  is the usual
commutator of derivations:
\begin {equation}
[D_1,D_2]=D_1D_2-D_2D_1
\end {equation}

Let us give another (equivalent) description of this Lie algebra. Consider  the
convolution product $f \star g$, of two elements $f,g \in {\cal A}_p^{*}$ (the
topological (weak) dual of ${\cal A}_p$):
\begin {eqnarray}
f \star g (X^{\omega_1 ... \omega_r}) &=& (f \otimes g) \circ
\Delta (X^{\omega_1 ... \omega_r}) \nonumber \\
  &=& \sum_{i=0}^r f(X^{\omega_1 ... \omega_i}).g(X^{\omega_{i+1} ...
\omega_r})
\end {eqnarray}

 We need the following Lemma (see [Ab]):

\subsubsection {Lemma}

 $({\cal A}_p^{*},\star)$  is a topological {\bf k}-algebra, isomorphic (resp.,
antiisomorphic) to the topological algebra $End^{ll}({\cal A}_p)$ (resp.,
$End^{rl}({\cal A}_p)$) of all left (resp., right) invariant {\bf k}-linear
endomorphisms of ${\cal A}_p$ (i.e., {\bf k}-linear morphisms $\sigma:{\cal
A}_p \rightarrow {\cal A}_p$ that verify the left (resp., right) invariance
condition:  \begin{equation}
\Delta \circ \sigma = (1 \otimes \sigma) \circ \Delta
\end {equation}
(resp., $\Delta \circ \sigma = ( \sigma \otimes 1) \circ \Delta$), and endowed
with the topology of pointwise convergence. Moreover, each element of
$End^{ll}({\cal A}_p)$ commutes with each element of $End^{rl}({\cal A}_p)$
\medskip

{\underline {Proof}}...

{\small It's a standard fact that $({\cal A}_p^{*},\star)$  is a
topological {\bf k}-algebra.

Define now {\bf k}-linear maps $\Phi: End^{ll}({\cal A}_p) \rightarrow {\cal
A}_p^{*}$, $\Psi: {\cal A}_p^{*} \rightarrow End^{ll}({\cal A}_p)$, and
$\Lambda: {\cal A}_p^{*} \rightarrow End^{rl}({\cal A}_p)$ respectivelly by:
\begin {equation}
\Phi:  \sigma \rightarrow \Phi(\sigma) \equiv f_{\sigma} \equiv \epsilon
\circ \sigma
\end {equation}

\begin {equation}
\Psi:  f \rightarrow \Psi(f) \equiv {\sigma}_f \equiv  (1 \otimes f) \circ
\Delta
\end {equation}
and
\begin {equation}
\Lambda:  f \rightarrow \Lambda(f) \equiv {\rho}_f \equiv  (f \otimes 1) \circ
\Delta
\end {equation}
Let us verify, for example, that $\Psi$ is well defined and it's an algebra
morphism. In fact:
\begin {eqnarray}
\big(1 \otimes \Psi(f)\big)\Delta &=& \big(1 \otimes (1 \otimes f) \circ
\Delta \big)\Delta \nonumber \\
&=&(1\otimes 1 \otimes f)(1 \otimes \Delta)\Delta \nonumber \\
&=& (1\otimes 1 \otimes f)( \Delta \otimes 1)\Delta \nonumber \\
&=& \Delta (1 \otimes f)\Delta = \Delta \Psi(f)
\end {eqnarray}
which proves that $\Psi(f)$ is left invariant. On the other hand, composing
with $1 \otimes g$, $g \in {\cal A}_p^{*}$, we obtain:

\begin {eqnarray}
\Psi(g)\Psi(f)&=&(1 \otimes g)\Delta \Psi(f) \nonumber \\
&=& (1 \otimes g)(1
\otimes \Psi(f))\Delta   \nonumber \\
&=& (1 \otimes g \Psi(f))\Delta \nonumber \\
&=& (1 \otimes g \star f) \Delta \nonumber \\
&=& \Psi (g \star f)
\end {eqnarray}

The rest of the proof follows directly from the definitions , QED.}
\medskip
\medskip

\subsubsection {Example}

{\small Let $\alpha \in {\bf L}{\cal M}_p$ and ${\delta}_{\alpha} \in {\cal
A}_p^{*}$ as in (61). Then $\Psi_{{\delta}_{\alpha}}$ is the automorphism
$X^{\bf u} \rightarrow {\alpha} \cdot X^{\bf u} $  corresponding to the
action of $\alpha$, on ${\bf L}{\cal M}_p$ from the right.

In fact, the right action of ${\bf L}{\cal M}_p$ on itself, through right
translations $r_{\alpha}: \beta \mapsto \beta.\alpha$, induces a left action
of ${\bf L}{\cal M}_p$ on ${\cal A}_p$ by:
\begin {equation}
(\alpha \cdot X^{\bf u})(\beta) \equiv X^{\bf u}(\beta.\alpha)
\end {equation}

By the identification $\alpha \rightarrow {\delta}_{\alpha}$, we can write the
RHS of the above equation in the form:
\begin {eqnarray}
X^{\bf u}(\beta.\alpha)&=& {\delta}_{\beta.\alpha}(X^{\bf u})={\delta}_{\beta}
\star {\delta}_{\beta}(X^{\bf u}) \nonumber \\
&=&{\delta}_{\beta}\big((1 \otimes {\delta}_{\alpha})\Delta X^{\bf
u}\big)={\delta}_{\beta}\big(\Psi_{{\delta}_{\alpha}}(X^{\bf u}\big)
\end {eqnarray}
while the LHS is simply ${\delta}_{\beta}(\alpha \cdot X^{\bf u})$, which
allows the above mentioned identification $\Psi_{{\delta}_{\alpha}} \simeq
 \alpha \cdot X^{\bf
u}$.

In the same way we can prove that $\Lambda_{{\delta}_{\alpha}}$ is the
automorphism $X^{\bf u} \rightarrow
X^{\bf u} \cdot {\alpha} $,  corresponding to the
action of $\alpha$, on ${\bf L}{\cal M}_p$ from the
left.}
\medskip
\medskip

Now, if moreover $\sigma=D$ is a left invariant derivation, then the
corresponding $\Phi(D)=f_D=\epsilon \circ D \in {\cal A}_p^{*}$ verifies:
\begin {eqnarray}
f_D(X^{\bf u}X^{\bf v})&=&\epsilon D(X^{\bf u}X^{\bf v}) \nonumber \\
&=&\epsilon(X^{\bf u}DX^{\bf v} + DX^{\bf u}X^{\bf v}) \nonumber \\
&=&\epsilon(X^{\bf u})f_D(X^{\bf v}) + f_D(X^{\bf u})\epsilon(X^{\bf v})
\end {eqnarray}
and we see that the Lie algebra $\widetilde {l{\cal {M}}_p}$ is isomorphic,
as {\bf k}-linear space, to the subspace of ${\cal A}_p^{*}$ consisting of
the so called {\it point derivations} at $\epsilon$, i.e.:
\begin {equation}
\widetilde {l{\cal {M}}_p} \cong \{\delta \in {\cal A}_p^{*}: \ \
\delta(X^{\bf u}X^{\bf v})=\epsilon(X^{\bf u})\delta (X^{\bf v}) +
\delta(X^{\bf u})\epsilon(X^{\bf v}) \}. \end {equation}

This {\bf k}-linear space of point derivations at $\epsilon$ is called the
{\it Tangent Space}, at $\epsilon$, to the group  ${\widetilde {{\bf L}{\cal
M}_p}}$ and is denoted by $T_{\epsilon}{\widetilde {{\bf L}{\cal
M}_p}}$. The reason for this terminology can be explained in the following
more familiar terms. Let $\tilde \alpha_t$ be a curve of generalized loops
such that:
\begin {eqnarray}
{\tilde \alpha_0}&=&\epsilon\\
\lim_{t \rightarrow 0}{\tilde \alpha_t}&=&\epsilon\\
\lim_{t \rightarrow 0}\frac {{\tilde \alpha_t}-\epsilon}{t}&=&\delta \ \ \in
{\cal A}_p^{*} \end {eqnarray}
where the limits in (83) and (84) are taken in the weak sense (for example,
(83) means that $\lim_{t \rightarrow 0}{\tilde
\alpha_t}(X^{\bf u})=\epsilon(X^{\bf u}), \forall {\bf u} \in Sh$, and an
analogous condition for (84)). Then:
\begin {eqnarray}
\delta(X^{\bf u}X^{\bf v})&=&\lim_{t \rightarrow 0}\frac {{\tilde
\alpha_t}(X^{\bf u}X^{\bf v}) -\epsilon(X^{\bf u}X^{\bf v})}{t} \nonumber \\
&=&\lim_{t \rightarrow 0}\Big({\tilde \alpha_t}(X^{\bf u})\frac
{ {\tilde \alpha_t}(X^{\bf v}) - \epsilon(X^{\bf v})}{t} + \frac {{\tilde
\alpha_t}(X^{\bf u})-\epsilon(X^{\bf u})}{t} \epsilon(X^{\bf v}) \Big)
\nonumber
\\ &=& \epsilon(X^{\bf u})\delta (X^{\bf v}) + \delta(X^{\bf u})\epsilon(X^{\bf
v})  \end {eqnarray}

Thus, we have a {\bf k}-linear isomorphism:
\begin {equation}
T_{\epsilon}{\widetilde {{\bf L}{\cal
M}_p}} \cong \widetilde {l{\cal {M}}_p}
\end {equation}
given by $\delta \rightarrow D_{\delta}=(1 \otimes  \delta) \circ \Delta$.
We endow $T_{\epsilon}{\widetilde {{\bf L}{\cal
M}_p}}$ with a Lie brackett, by defining:
\begin {equation}
[\delta,\eta] \equiv \epsilon \circ [D_{\delta},D_{\eta}].
\end {equation}

Using lemma 3.4.2, we see that:
\begin {eqnarray}
[\delta,\eta ] &=& {\epsilon}  \circ [D_{\delta},D_{\eta}]
\nonumber \\
 &=& {\epsilon} (D_{\delta} D_{\eta} - D_{\eta} D_{\delta}) \nonumber \\
{} &=& \Phi (D_{\delta}D_{\eta}-D_{\eta} D_{\delta}) \nonumber \\
{} &=& \Phi (D_{\delta}) \star \Phi (D_{\eta}) - \Phi (D_{\eta})
\star \Phi (D_{\delta} )\nonumber \\
&=&\delta \star \eta - \eta \star \delta
\end {eqnarray}

Note that any point derivation $\delta$, at $\epsilon$, verifies:
\begin {equation}
\delta(X^{\omega_1 ... \omega_r}X^{\omega_{r+1} ... \omega_{r+s}})=0
\end {equation}
$\forall r \geq 1, \forall s \geq 1$, and from this we can deduce that:
\begin {equation}
{\delta}^{n}(X^{\omega_1 ... \omega_r}) =0 \ \ \ \ \ \ \forall n>r\geq 0
\end {equation}
where ${\delta}^{n} \equiv {\delta}^{n} \star {\delta},\ \  \forall n \geq 1$.

Now, for each $\delta \in T_{\epsilon}{\widetilde {{\bf L}{\cal
M}_p}} \cong \widetilde {l{\cal {M}}_p}$, define $\exp \delta$ by:
\begin {equation}
\exp {\delta} \equiv {\epsilon} + {\sum}_{n \geq 1} \frac {{\delta}^{n}}{n!}
\end {equation}
where, as always, this means that, for each $X^{\omega_1 ... \omega_r}$,
$\exp {\delta}(X^{\omega_1 ...\omega_r})$ is defined by:
\begin {equation}
\Big( {\epsilon} + {\sum}_{n \geq 1} \frac {{\delta}^{n}}{n!} \Big)
( X^{\omega_1... \omega_r})
\end{equation}
if, of course, this series converges.

Now, it follows from (90) that  the series (92) is in fact a finite sum, and
so $\exp {\delta}$ is well defined, in the above sense. Moreover, we can prove
that $\exp {\delta}$ is a generalized loop (definition 3.3.1).

Converselly, given ${\tilde {\alpha}} \in {\widetilde {{\bf L}{\cal M}_p}}
$, we define:
\begin {equation}
\log {\tilde {\alpha}} \equiv {\sum}_{n \geq 1} \frac {(-1)^{n-1}}{n}({\tilde
{\alpha}}-{\epsilon})^{n}
\end {equation}
where $({\tilde
{\alpha}}-{\epsilon})^{n} \equiv ({\tilde
{\alpha}}-{\epsilon})^{n-1} \star ({\tilde
{\alpha}}-{\epsilon}), \ \ \forall n \geq 1$.

Since, $({\tilde {\alpha}}-{\epsilon})^{n}( X^{\omega_1... \omega_r})=0,\ \
\forall n>r \geq 0$, $\log {\tilde {\alpha}}$ is also a well defined element in
the above sense, which moreover, belongs to $T_{\epsilon}{\widetilde {{\bf
 L}{\cal
M}_p}} \cong \widetilde {l{\cal {M}}_p}$. By the calculus of formal power
series, we know that:
\begin {eqnarray*}
\exp (k \log {\tilde {\alpha}}) &=& {\tilde {\alpha}}^k \ \ \ \ \ \forall k \in
Z \\
\log (\exp {\delta}) &=& {\delta}
\end {eqnarray*}

Let us define, for each $t \in {\bf k}$:
\begin {equation}
{\tilde {\alpha}}^t \equiv \exp (t \log {\tilde {\alpha}})
\end{equation}

Then we can easilly prove that $t \mapsto {\tilde {\alpha}}^t$ is a
one-parameter subgroup of ${\widetilde {{\bf L}{\cal M}_p}}$,
generated by $\log {\tilde {\alpha}}$, i.e.:
\begin {eqnarray*}
{\tilde {\alpha}}^0 &=& \epsilon \\
{\tilde {\alpha}}^t \star {\tilde {\alpha}}^s &=& {\tilde {\alpha}}^{t+s} \\
\lim _{t \rightarrow 0} \frac {{\tilde {\alpha}}^t - \epsilon}{t} &=& \log
{\tilde {\alpha}}
\end {eqnarray*}
this last limit in the above (weak) sense. Similar results seem to be obtained
in [BGG], using however a quite different formalism.

 \medskip

\section {Loop Calculus. Endpoint and Area  Derivative Operators}

In this section we try to develope a rigorous mathemathical {\it Loop Calculus}
which can be used to formalize early heuristic ideas mainly due to Gambini and
Trias (see [GT1],[GT2],[GT3]), and also Makeenko, Migdal (see [MM]).

\subsection {Endpoint Derivatives}

 Consider a path $\gamma \in {\cal PM}$, and a tangent vector field
 $V \in {\cal XU}$, defined in a neighbourhood $\cal U$ of
$q=\gamma(1)$ and such that $V(\gamma(1)) = v \in T_{\gamma (1)}{\cal
M}$ .
 Let us put $\eta^V_{ s} = {\Phi}^{V}( s)(q)$   for the
integral curve of $V$, starting at $q=\gamma (1)$, at $ s = 0$,
with velocity $v$. Finally put
$\gamma_{ s}=\gamma.\eta^V_{ s}$ and $q_{ s} =
\gamma_{ s}(1)$ (see figure 1).

\medskip
 \begin{figure}[htb]
\vspace{1in}
\caption{ $ \gamma_{ s}=\gamma.\eta^V_{ s}$ and $q_{ s} =
\gamma_{ s}(1) $}.
\end{figure}
\medskip

We use these notations in the following:

\subsubsection {Definition}

Let $\Psi$ be a path functional on ${\cal PM}$, with values in $R$ (resp.,
$C; gl(m)$). We define the {\it Terminal Covariant Endpoint
Derivative} $\nabla^T_{V}(q_{ s}) \Psi (\gamma)$, of
$\Psi$, at $\gamma$, in the direction of  $V$, as the limit:

\begin {equation}
\nabla^T_{V}(q_{ s}) \Psi (\gamma) = \lim_{h \rightarrow 0} \frac
{\Psi( {\gamma}_{ s +h}) - \Psi(\gamma_{ s})}{h}.
\end {equation}

\subsubsection {Note}

{\small  With the obvious modifications (in particular
$\gamma_{ s}$ now means
$\gamma_{ s}=(\eta^V_{ s})^{-1}.{\gamma}$), we also
define the {\it Initial Covariant Endpoint Derivative}
$\nabla^I_{V}(q_{ s}) \Psi (\gamma)$, as the limit given by the
same formal expression. Hereafter, to simplify the discussion, we treat
only the {\it Terminal Endpoint} case. Similar formulas can be obtained
for the {\it Initial Endpoint} case, with the obvious modifications.}

\medskip

\medskip

 Note that the limit (95) (if it exists) defines
$\nabla^T_{V}(q_{ s}) \Psi (\gamma)$ "near" the endpoint
$q={\gamma}(1)$, more preciselly at the point
$q_{ s}=\gamma_{ s}(1)$, and obviously depend on the vector
field $V$. When $ s =0$, we define:

\subsubsection {Definition}

Let $\Psi$ be a path functional on ${\cal PM}$, with values in $R$ (resp.,
$C; gl(m)$). We define the {\it Terminal Endpoint Derivative}
$\partial^T_{v} \Psi (\gamma)$,  of
$\Psi$, at $\gamma$, in the direction of  $v \in T_{\gamma(1)}{\cal
M}$, as the limit:

\begin {equation}
\partial^T_{v} \Psi (\gamma) = \lim_{h \rightarrow 0} \frac
{\Psi( {\gamma}_{h}) - \Psi(\gamma)}{h}
\end {equation}
provided this limit exists independently of the choice of the vector
field $V \in {\cal XM}$, such that $V(\gamma(1)) = v$.

\medskip
\medskip

 As a simple example, let us take a smooth function $f \in
C^{\infty}{\cal M}$, and define a path functional $\Psi_f$, by
$\Psi_f(\gamma)=f(\gamma(1))$. Then it's easy to see that:

\begin {equation}
\nabla^T_{V}(q_{ s}) \Psi_{f} (\gamma) = V.f(q_{ s}) =
df(V_{q_{ s}}).
\end {equation}

Moreover,

\begin {equation}
\partial^T_{v} \Psi_{f} (\gamma) = V.f(\gamma (1)) =
df(v)
\end {equation}
 which depends only on the vector $v$, and not of the particular
extension $V$.

\subsubsection {Definition}

Let $\Psi$ be a path functional and $f  \in
C^{\infty}{\cal M}$. We  define the {\it Marked Path Functional} $f
\odot \Psi$, by:
\begin {equation}
(f \odot \Psi)(\gamma)=f(\gamma(1)) \Psi (\gamma).
\end {equation}

The following lemma is proved as usual, following the definitions.

\subsubsection {Lemma. Leibniz Rule}

Assume that $\Psi$ is a path functional for which the limit (95)
exists, and which verifies the "continuity condition" $\lim_{h
\rightarrow 0} \Psi({\gamma_{ s +h}})=
\Psi({\gamma_{ s}}), \forall  s \geq 0$ . Then we have the
following {\it Leibniz Rule}:

\begin {eqnarray}
\nabla^T_{V}(q_{ s}) \big( f \odot \Psi \big) (\gamma) &=&
V.f(q_{ s}) \Psi (\gamma_{ s}) + f (q_{ s})
\nabla^T_{V}(q_{ s}) \Psi (\gamma)  \nonumber \\ &=&
\nabla^T_{V}(q_{ s}) f (\gamma) \Psi (\gamma_{ s}) +
f(\gamma_{ s}) \nabla^T_{V}(q_{ s}) \Psi (\gamma) \end
{eqnarray} where we have put $q_{ s}=\gamma_{ s}(1)$, and
used the notation $f$ for $\Psi_f$.

In particular, if $\partial^T_{v} \Psi (\gamma)$ exists in the
sense of definition 4.1.3, we have at the terminal endpoint
$q=\gamma(1)$:
\begin {equation}
\partial^T_{v} \big( f \odot \Psi \big) (\gamma) = \partial^T_{v} f
(\gamma) \Psi (\gamma) + f(\gamma) \partial^T_{V} \Psi (\gamma)
\end {equation}
 which depends only on the vector $v$, and not of the particular
extension $V$.

\medskip
\medskip

Our aim now, is to prove that the path functionals
$X^{\omega_1...\omega_r}$, defined by
(2), have well defined endpoint derivatives in the above sense. For
this, we need the following lemma which is easilly proved in local
coordinates.

\subsubsection {Lemma}

Let ${\eta_{ s}}={\eta^V_{ s}}$. Then:
\begin {eqnarray}
\lim_{ s \rightarrow 0} \frac
{\int_{\eta_{ s}} \omega}{ s} &=& \omega
(v) \\ \lim_{ s \rightarrow 0} \frac {\int_{\eta_{ s}}
\omega_1 ... \omega_r}{ s} &=& 0 \ \ \ \  \forall r \geq 2
\end {eqnarray}
where $\omega, \omega_1,...,\omega_r \in {\bigwedge}^{1}{\cal M}$ (resp.,
 $\omega, \omega_1,...,\omega_r \in {\bigwedge}^{1}{\cal M}
\otimes gl(m)$).

\medskip
\medskip

 Then, using the above lemma, Proposition 2.4 and induction on $r \geq
1$, we  compute that:
\medskip
\medskip
\begin {equation}
 \nabla^T_{V}(q_{ s}) X^{\omega_1...\omega_r}(\gamma) =
X^{\omega_1...\omega_{r-1}}({\gamma}_{ s}). \omega_r
(V_{q_{ s}}) \ \ \ \forall  r \geq 1
\end {equation}
and:
\medskip
\medskip
\begin {equation}
 \partial^T_{v} X^{\omega_1...\omega_r}(\gamma) =
X^{\omega_1...\omega_{r-1}}(\gamma). \omega_r
(v) \ \ \ \forall  r \geq 1
\end {equation}

\medskip
\medskip

In the same way (see Note 4.1.2), we can compute that:
\begin {equation}
 \nabla^I_{V}(q_{ s}) X^{\omega_1...\omega_r}(\gamma) =  -\omega_1
(V_{q_{ s}}).
X^{\omega_2...\omega_{r}}({\gamma}_{ s})   \ \ \ \forall  r \geq 1
\end {equation}
and:
\begin {equation}
 \partial^I_{v} X^{\omega_1...\omega_r}(\gamma) = -\omega_1 (v).
X^{\omega_2...\omega_r} \ \   \forall  r \geq 1
\end {equation}
where $ \omega_1,...,\omega_r \in {\bigwedge}^{1}{\cal M}$ (resp.,
 $ \omega_1,...,\omega_r \in {\bigwedge}^{1}{\cal M}
\otimes gl(m)$).

\medskip
\medskip

 \subsubsection {Note}

 {\small It's easy to see that $\partial^T_{v}$ (resp.,
$\partial^I_{v}$)  is a derivation in the algebra of iterated
integrals. In fact, this follows from the fact that $\int$ is an
algebra homomorphism and $\partial^T_{v}:\omega_1...\omega_r \mapsto
  \omega_1...\omega_{r-1}.(\omega_r(v))$ (resp.,
$\partial^I_{v}:\omega_1...\omega_r \mapsto -\omega_1(v).
\omega_2...\omega_{r}$) is a derivation in $Sh({\cal M})$ (we put
also $\partial^T_{v}$ (resp., $\partial^I_{v}$): $1 \mapsto 0$, when
$r=0$).

In this spirit, we can compute the (algebraic)  commutator of two
endpoint derivatives (considered as derivations on the algebra of
iterated integrals), applied to a function
$X^{\omega_1...\omega_r}$. It is given by:

\begin {equation}
[\partial^T_{u},\partial^T_{v}] X^{\omega_1...\omega_r}(\gamma) =
X^{\omega_1...\omega_{r-2}}(\gamma). (\omega_{r-1} \wedge {\omega_r}) (u \wedge
v).   \end {equation}}

\medskip
\medskip

Note that $\nabla^T_{V}(q_{ s}) X^{\omega_1...\omega_r}(\gamma)$,
given by (104), is a Marked Path Functional, in the sense of definition
4.1.4 ($f$ is the function $\omega_r(V)$). So, if $U,V$ are two vector
fields locally defined around $q=\gamma(1)$, we can apply Leibniz
rule (101) and the well known formula $d\omega
(U,V)=U.\omega(V)-V.\omega(U)-\omega([U,V])$, to compute that, at
$q$:

\begin {eqnarray}
\big[\nabla^T_{U}(q),\nabla^T_{V}(q)\big]
X^{\omega_1...\omega_r}(\gamma) &=&
X^{\omega_1...\omega_{r-1}}(\gamma). d{\omega_r}(u \wedge v)\nonumber \\
      &  & \ \ \  +   X^{\omega_1...\omega_{r-2}}(\gamma).  ({\omega_{r-1}}
\wedge {\omega_{r}})(u \wedge v).
\nonumber \\
& &
\end {eqnarray}

As the RHS of the above formula, depends only on the vectors $u,v$,
and not of the particular extensions $U,V$, we write the LHS as $
\big[\nabla^T_{u},\nabla^T_{v}\big] X^{\omega_1...\omega_r}(\gamma)$.

\medskip
\medskip

For the parallel transport path functional $U: {\cal PM} \rightarrow
Gl(p)$, given by the series (27), we compute that:

\begin {equation}
\partial^T_{v}U_{\gamma} =   U_{\gamma}.\omega (v),
\end {equation}

\begin {equation}
\partial^I_{v}U_{\gamma} = -\omega (v). U_{\gamma},
\end {equation}
while, for the commutator:
\begin {eqnarray}
\big[\nabla^T_{u},\nabla^T_{v}\big] U_{\gamma} &=&  U_{\gamma}. (d{\omega} +
\omega \wedge \omega)(u \wedge v)\nonumber \\
                       &=&   U_{\gamma}. \Omega (u \wedge v)
\end {eqnarray}
where $\Omega$ is the curvature of the connection $\omega$.
\medskip
\medskip

 Let us give a last example, to finish this section. Given a path $\lambda \in
{\cal PM}_p$ and a function $f \in C^{\infty}{\cal M}$ (resp., $ \in
C^{\infty}{\cal M} \otimes gl(m)$), we define a (marked) path functional
through: \begin {equation}
Z_{(i)}^{\omega_1...\omega_r}(\lambda;f) \equiv
X^{\omega_1...\omega_i}(\lambda)f(\lambda(1))
X^{\omega_{i+1}...\omega_r}({\lambda}^{-1})
\end{equation}
where $ \omega_1,...,\omega_r \in {\bigwedge}^{1}{\cal M}$ (resp.,
 $ \omega_1,...,\omega_r \in {\bigwedge}^{1}{\cal M}
\otimes gl(m)$).

Using Leibniz rule, we compute that:
\begin {eqnarray}
\nabla_v^{T}Z_{(i)}^{\omega_1...\omega_r}(\lambda;f)
&=&X^{\omega_1...\omega_i}(\lambda).df_q(v)
X^{\omega_{i+1}...\omega_r}({\lambda}^{-1}) \nonumber \\
&& + X^{\omega_1...\omega_{i-1}}(\lambda).\omega_i(v).f(q).
X^{\omega_{i+1}...\omega_r}({\lambda}^{-1}) \nonumber \\
&& + X^{\omega_1...\omega_i}(\lambda).f(q).\omega_{i+1}(v).
X^{\omega_{i+2}...\omega_r}({\lambda}^{-1})
\end{eqnarray}
Where $q=\lambda(1)$.

Finally, let us define, for a connection 1-form $\omega$, a (marked) path
functional $\Psi$ through:
\begin {equation}
\Psi(\lambda;f) \equiv U_{\lambda}.f(q).U_{\lambda^{-1}}
\end{equation}
where $q=\lambda(1)$, $U$ is the parallel transport operator of the connection
$\omega$ and $f \in C^{\infty}{\cal M} \otimes gl(m)$. Then, using (114), we
compute that:
\begin {eqnarray}
\nabla_v^T \Psi(\lambda;f)
&=& U_{\lambda}.\big( df_q(v)+[{\omega},f](v) \big).U_{\lambda^{-1}}\nonumber
\\
&=& U_{\lambda}.D^{\omega}_q f(v).U_{\lambda^{-1}}
\end{eqnarray}
where $D^{\omega}_q f(v) \equiv df_q(v)+[{\omega},f](v)$ denotes the covariant
derivative of $f$. This is the reason why we call the operator $\nabla^T_v$,
 Terminal Endpoint {\it Covariant} Derivative.

\medskip
\medskip

 \subsection {Area Derivative}

\subsubsection {Definition and Main Properties}

 Consider a loop $\gamma \in {\cal LM}_p$  , a point $q \in {\cal M}$
and a path $\lambda \in {\cal PM}_p$, going from $p$ to
$q={\lambda}(1)$.

 Given an ordered pair $(u,v)$ of tangent vectors $u,v \in
T_q{\cal M}$, we extend them by two commuting vector
fields $U,V \in {\cal XU}$, defined in a small
neighbourhood $\cal U$ of $q={\lambda}(1)$ (this is always
possible). Then, we consider the "small" loop
$\Box^{(U,V)}_{t}$, based at $q$, defined by:

\begin {equation}
\Box^{(U,V)}_{t} = \Phi^{V}(-t) \Phi^{U}(-t)
\Phi^{V}(t) \Phi^{U}(t)(q)
\end {equation}
where $\Phi^{U}$ (resp., $\Phi^{V}$) denotes the local flow of $U$
(resp.,$V$).

\medskip
Denote by $\lambda_t$ the ($t$-dependent) loop
$\lambda.\Box^{(U,V)}_{t}.{\lambda}^{-1}$ (see figure 2).

\begin{figure}[htb]
\vspace{1in}
\caption{ $\lambda_t \equiv \lambda.\Box^{(U,V)}_{t}.{\lambda}^{-1}
 $}.
\end{figure}

\medskip
Note that $\lim_{t \rightarrow 0}{\lambda}_t =  \epsilon$,
where $\epsilon$ denotes the unity in the  group ${\bf
L}{\cal M}_p$ of (equivalence classes) of loops based at
$p \in {\cal M}$ (i.e., $\epsilon \equiv
[p]$, the equivalence class of the trivial loop reduced to
the point $p$), and the limit is taken in weak sense,
i.e.:
\begin {equation}
\lim_{t \rightarrow 0}{\lambda_t}(X^{\bf
u})=\lim_{t \rightarrow 0} X^{\bf
u}({\lambda_t})=\epsilon(X^{\bf u})
\end {equation}
$\forall {\bf u} \in Sh({\cal M})$ (see section 3.3). So,
$\lambda_t.\gamma$ represents a "small" deformation of
the loop $\gamma$, in the  topology of ${\bf
L}{\cal M}_p$, defined in section 3.3.

\medskip

 With all these notations we now formulate the following:

\subsubsection {Definition}
Given a loop functional $\Psi$ on ${\bf L}{\cal M}_p$, with
values in $R$ (resp., $C; gl(m)$), we define its {\it Area
Derivative}, denoted by $\triangle_{\lambda;(u,v)}(q).\Psi
(\gamma)$, as the limit:

\begin {equation}
\triangle_{\lambda;(u,v)}(q) \Psi (\gamma) = \lim_{t \rightarrow 0}
\frac {\Psi(\lambda_t.\gamma  ) -
\Psi(\gamma)}{t^2} \end {equation}
provided this limit exists independently of the choice of the vector
fields $U,V \in {\cal XU}$, considered above.

\medskip
\medskip

We want to prove that every function
$X^{\omega_1...\omega_r}:{\bf L}{\cal M}_p\rightarrow R$ (resp., $C; gl(m)$)
admits a well defined Area derivative, in the above sense. For this,
we first state a lemma which is, once more, easilly proved in local
coordinates:

\subsubsection {Lemma}

Denote simply by ${\Box}_{t}$, the above mentioned loop
${\Box}^{(U,V)}_{t}$. Then:
\begin {eqnarray}
\lim_{t \rightarrow 0} \frac {\int_{{\Box}_{t}}
\omega}{t^2} &=& d\omega (u \wedge v) \\
\lim_{t \rightarrow 0} \frac {\int_{{\Box}_{t}}
\omega_1 \omega_2}{t^2} &=& (\omega_1 \wedge \omega_2)(u
\wedge v)\\ \lim_{t \rightarrow 0} \frac
{\int_{{\Box}_{t}} \omega_1 ...
\omega_r}{t^2} &=& 0, \ \ \ \  \forall r \geq 3
\end {eqnarray}
 where $\omega, \omega_1,...,\omega_r \in {\bigwedge}^{1}{\cal M}$ (resp.,
 $\omega, \omega_1,...,\omega_r  \in C; {\bigwedge}^{1}{\cal M} \otimes
gl(m)$).

\medskip
\medskip

To give a compact form  for the expression of the area derivative of the
path functionals $X^{\omega_1...\omega_r}$, we introduce some more
notations.

First, for $u \wedge v \in {\bigwedge}^{2}T_q{\cal M}$, we define a
derivation $D_{u \wedge v}(q)$, in the algebra of iterated integrals, by the
formula:

\begin {equation}
D_{u \wedge v}(q) X^{\omega_1...\omega_r}  =
X^{\omega_1...\omega_{r-1}}. d\omega_r(u \wedge v)
\end {equation}

Second, recalling formula (108) for the (algebraic) commutator
$[\partial^T_{u},\partial^T_{v}]$ of two terminal endpoint derivatives at
$q$, we define a new derivation ${\cal D}_{u \wedge v}(q) $ by:
\begin {equation}
{\cal D}_{u \wedge v}(q) = D_{u \wedge v}(q) +
[\partial^T_{u},\partial^T_{v}]
\end {equation}

Let us   evaluate first, the area derivative of
$X^{\omega_1...\omega_r}$ at $\epsilon$.  Using lemma 4.2.3,
Proposition 2.4 and induction in $r$, we can prove  the following
lemma:

\medskip
\subsubsection {Lemma}

\begin {equation}
\triangle_{\lambda;(u,v)}(q) X^{\omega_1...\omega_r}(\epsilon)
 = \sum_{i=1}^r \Big( {\cal D}_{u
\wedge v}(q)X^{\omega_1...\omega_i}(\lambda) \Big) \Big(
X^{\omega_{i+1}...\omega_r} (\lambda^{-1}) \Big)
\end {equation}
\medskip
\medskip
 where $\omega_1,...,\omega_r \in
{\bigwedge}^{1}{\cal M}$ (resp., $\in C,  {\bigwedge}^{1}{\cal M} \otimes
gl(m)$).

Note that $\triangle_{\lambda;(u,v)}(q)
X^{\omega_1...\omega_r}(\epsilon)$ depends on the pair
$(u,v)$ through $u \wedge v$. So we write it in the form $
\triangle_{(\lambda;u \wedge v)}(q)
X^{\omega_1...\omega_r}(\epsilon)$.
\medskip

{\small For example:
\begin {eqnarray}
\triangle_{(\lambda;u \wedge v)}(q) X^{\omega}(\epsilon) &=& d\omega
(u \wedge v) \nonumber \\
\triangle_{(\lambda;u \wedge v)}(q)
X^{\omega_1 \omega_2}(\epsilon) &=&  d\omega_1 (u \wedge v)
 X^{\omega_2}(\lambda^{-1}) +
X^{\omega_1}(\lambda). d\omega_2 (u \wedge v) +  \nonumber \\
& & + (\omega_1 \wedge \omega_2)(u \wedge v) \nonumber \\
\end {eqnarray}
and, more generally:
\begin {eqnarray}
&&\triangle_{(\lambda;u \wedge v)}(q) X^{\omega_1...\omega_r}(\epsilon)=
\sum_{i=1}^r X^{\omega_1...\omega_{i-1}}(\lambda).d{\omega_i}(u \wedge v)
 X^{\omega_{i+1}...\omega_r}(\lambda^{-1}) + \nonumber \\
&& \ \ \ \ \ \ \ \ \ \ \  +
\sum_{i=2}^r X^{\omega_1...\omega_{i-2}}(\lambda).(\omega_{i-1} \wedge
{\omega_i})(u \wedge v). X^{\omega_{i+1}...\omega_r}(\lambda^{-1})
\end {eqnarray}}

Recall that $\lim_{t \rightarrow 0}{\lambda}_t = \epsilon$. We can
also prove that $\lim_{t \rightarrow 0}\frac{{\lambda}_t
-\epsilon}{t}= 0$, while $\lim_{t \rightarrow 0}\frac{{\lambda}_t
-\epsilon}{t^2}$ exists and is our area derivative . To be
more specific, let us define an operator ${\bf
{\delta}}_{(\lambda;u \wedge v)}$ in the algebra of iterated
integrals ${\cal A}_p$, through the formula:
\begin {eqnarray}
{\bf {\delta}}_{(\lambda;u \wedge v)} X^{\bf u} &\equiv&
\triangle_{(\lambda;u \wedge v)}(q) X^{\bf u}(\epsilon) \nonumber \\
&=& (\lambda \otimes \lambda) \big( ({\cal D}_{u \wedge v}(q) \otimes J)
\circ \Delta \big) X^{\bf u}
\end {eqnarray}
$\forall {\bf u} \in Sh$. Note that the last equality in (127), is
nothing else that (126),  written using the definitions from
section 3.

The above formula shows that ${\bf {\delta}}_{(\lambda;u \wedge
v)}:{\cal A}_p \rightarrow {\bf k}$ is linear and we can also prove
easilly that:
\begin {equation}
{\bf {\delta}}_{(\lambda;u \wedge v)} (X^{\bf u}X^{\bf v})={\bf
{\delta}}_{(\lambda;u \wedge v)} (X^{\bf u})\epsilon(X^{\bf v}) +
\epsilon (X^{\bf u}){\bf {\delta}}_{(\lambda;u \wedge v)} (X^{\bf
v})  \end {equation}
$\forall {\bf u,v} \in Sh$. So, acording to section 3.4,
${\bf{\delta}}_{(\lambda;u \wedge v)}$ is a point derivation at
$\epsilon$. We call it a tangent vector to the group  ${\bf
L}{\cal M}_p$, at $\epsilon$.
\medskip

 By definition, the Tangent Space $T_{\epsilon}{\bf
L}{\cal M}_p$, to the group  ${\bf
L}{\cal M}_p$, at $\epsilon$, is the {\bf k}-linear subspace of
${\cal A}_p^{*}$, generated by all the  ${\bf {\delta}}_{(\lambda;u
\wedge v)}$. We can give a geometrical interpretation to the adition
and scalar multiplication on  $T_{\epsilon}{\bf
L}{\cal M}_p$.

\medskip
\medskip

Consider now a loop $\gamma \in {\bf L}{\cal M}_p$ and let us compute
the area derivative $\triangle_{(\lambda;u \wedge v)}(q)
X^{\omega_1...\omega_r} (\gamma) $, acording to definition (see figure 3).

\begin{figure}[htb]
\vspace{1in}
\caption{ $ \triangle_{(\lambda;u \wedge v)}(q)
X^{\omega_1...\omega_r} (\gamma) $}.
\end{figure}

We have, after an easy computation, the following lemma:

\subsubsection {Lemma}
Let $\gamma \in {\bf L}{\cal M}_p$, $\lambda \in {\cal PM}_p$ and $u
\wedge v \in {\wedge}^2 T_{\lambda(1)}{\cal M}$. Then:

\begin {eqnarray}
\triangle_{(\lambda;u \wedge v)}(q) X^{\omega_1...\omega_r} (\gamma)
&=& \sum_{i=1}^{r} \triangle_{(\lambda;u \wedge v)}(q)
X^{\omega_1...\omega_i}(\epsilon)
X^{\omega_{i+1}...\omega_r}(\gamma) \nonumber \\
&=& \gamma \circ ({\bf {\delta}}_{(\lambda;u
\wedge v)} \otimes 1) \circ \Delta ( X^{\omega_1...\omega_r})
\end {eqnarray}
where we have used, in the last equation, the formalism of section 3.

\medskip
\medskip

 Recall that $({\bf {\delta}}_{(\lambda;u
\wedge v)} \otimes 1) \circ \Delta $ is the right invariant
derivation on the algebra ${\cal A}_p$, associated to the tangent
vector ${\bf {\delta}}_{(\lambda;u \wedge v)}$ (see section 3.4). So, it
is natural to call the map ${\Lambda}^R_{(\lambda;u \wedge v)}:{\bf
L}{\cal M}_p \rightarrow {\cal A}_p^{*}$, given by:
\begin {equation}
{\Lambda}^R_{(\lambda;u \wedge v)}(\gamma) \equiv \gamma
\circ ({\bf {\delta}}_{(\lambda;u
\wedge v)} \otimes 1) \circ \Delta
\end {equation}
the {\it Right Invariant "Vector Field"}, on ${\bf
L}{\cal M}_p$, determined by ${\bf{\delta}}_{(\lambda;u \wedge v)}$.
\medskip
\medskip

In the special case in which $\lambda=\epsilon$, we call
$\triangle_{(\epsilon;u \wedge v)}(p)$ the {\it Initial Endpoint
Area Derivative}  and we denote it by $\triangle^I_{(\epsilon;u \wedge
v)}(p)$ (see figure 4).

\begin{figure}[htb]
\vspace{1in}
\caption{ $ \triangle^I_{(\epsilon;u \wedge
v)}(p) $}.
\end{figure}

Then, the above formula reduces to:
\begin {eqnarray}
\triangle^I_{(\epsilon;u \wedge v)}(p) X^{\omega_1...\omega_r}(\gamma)
 &=&  d{\omega_1}(u \wedge v). X^{\omega_2...\omega_{r}}(\gamma)
\nonumber \\
      &  & \ \ \  + ({\omega_{1}} \wedge
{\omega_{2}})(u \wedge v). X^{\omega_3...\omega_{r}}(\gamma).
\end {eqnarray}

\medskip
\medskip

Consider now the  case in which $\lambda=\gamma.\eta$, $\gamma
\in {\bf L}{\cal M}_p$, $\eta \in {\cal PM}_p$, and $u \wedge v \in
{\wedge}^{2}T_{\eta(1)}{\cal M}$.

Then $\lambda_t.\gamma \equiv
(\lambda.\Box^{(U,V)}_{t}.{\lambda}^{-1}).\gamma =
 \gamma.\eta.\Box^{(U,V)}_{t}.{ \gamma.\eta}^{-1}.\gamma =
\gamma.(\eta.\Box^{(U,V)}_{t}.\eta^{-1}) \equiv \gamma.\eta_t$ (see figure 5).

\begin{figure}[htb]
\vspace{1in}
\caption{ $ \triangle^E_{(\eta;u \wedge v)}(q) $}
\end{figure}

In this case, we call the corresponding area derivative {\it
 Endpoint Area Derivative} and we denote it by
$\triangle^E_{(\eta;u \wedge v)}(q)$. We compute that:
\begin {eqnarray}
\triangle^E_{(\eta;u \wedge v)}(q) X^{\omega_1...\omega_r} (\gamma)
&=&\sum_{i=1}^{r}
X^{\omega_{1}...\omega_i}(\gamma)\triangle_{(\eta;u \wedge v)}(q)
X^{\omega_{i+1}...\omega_r}(\epsilon) \nonumber
  \\ &=& \gamma \circ (1 \otimes {\bf
{\delta}}_{(\eta;u \wedge v)} ) \circ \Delta (
X^{\omega_1...\omega_r})
\end {eqnarray}

As before, since $(1 \otimes {\bf{\delta}}_{(\eta;u \wedge v)}
) \circ \Delta$ is the left invariant derivation associated to
${\bf{\delta}}_{(\eta;u \wedge v)}$, it
is natural to call the map ${\Lambda}^L_{(\eta;u \wedge v)}:{\bf
L}{\cal M}_p \rightarrow {\cal A}_p^{*}$, given by:
\begin {equation}
{\Lambda}^L_{(\eta;u \wedge v)}(\gamma) \equiv \gamma
\circ (1 \otimes{\bf {\delta}}_{(\eta;u
\wedge v)} ) \circ \Delta
\end {equation}
the {\it Left Invariant "Vector Field"}, on ${\bf
L}{\cal M}_p$, determined by ${\bf
{\delta}}_{(\eta;u \wedge v)}$.

Note the particular case $\eta=\epsilon$ (see figure 6), in which the above
formula reduces to:

\begin {eqnarray}
\triangle^E_{(\epsilon;u \wedge v)}(p) X^{\omega_1...\omega_r}(\gamma)
 &=& {\cal D}_{u \wedge v}(p) X^{\omega_1...\omega_r} \nonumber \\
&=&  X^{\omega_1...\omega_{r-1}}(\gamma). d{\omega_r}(u \wedge v) \nonumber\\
      &  & \ \ \  +   X^{\omega_1...\omega_{r-2}}(\gamma). ({\omega_{r-1}}
\wedge {\omega_{r}})(u \wedge v)
\end {eqnarray}

\begin{figure}[htb]
\vspace{1in}
\caption{ $ \triangle^E_{(\epsilon;u \wedge v)}(p) $}.
\end{figure}

Note that, in this case we have that (see (109)):
\begin {equation}
\triangle_{(\epsilon;u \wedge v)}(q) X^{\omega_1...\omega_r}(\gamma) =
 \big[\nabla^T_{u},\nabla^T_{v}\big] X^{\omega_1...\omega_r}(\gamma)
\end {equation}
which is an expected relation between the endpoint area derivative
and the commutator of two terminal endpoint covariant derivatives.

\medskip
\medskip
\medskip

For example, for a loop $\gamma$, based at $p \in {\cal M}$, the endpoint
area derivative of the holonomy $U_{\gamma}$   is given by:
\begin {eqnarray}
\triangle^E_{(\epsilon;u \wedge v)}(p)U_{\gamma} &=&   U_{\gamma}. (d{\omega} +
\omega \wedge \omega)(u \wedge v)\nonumber \\
                       &=&   U_{\gamma}. \Omega (u \wedge v)
\end {eqnarray}
where $\Omega$ is the curvature of the connection $\omega$. The endpoint
area derivative of the Wilson loop variable ${\cal W}$  is
given by:

\begin {eqnarray}
\triangle^E_{(\epsilon;u \wedge v)}(p){\cal W}({\gamma}) &=& Trace \big(
(d{\omega} + \omega \wedge \omega)(u \wedge v). U_{\gamma}
\big) \nonumber \\
                       &=& Trace \big( \Omega (u \wedge v).
U_{\gamma} \big).
\end {eqnarray}

 These formulas are known as Mandelstam formulas.

\medskip
\medskip

Another such relation is given by the so called {\it Bianchi
Identity}:

\subsubsection {Bianchi Identity}
\begin {equation}
\sum _{cycl\{u,v,w\}} \nabla^T_u(\lambda(1)) {\bf
{\delta}}_{(\lambda;v \wedge w)} = 0
\end {equation}

{\underline {Proof}}...

{\small It suffices to prove  that:
\begin {equation}
\sum _{cycl\{u,v,w\}} \nabla^T_u(\lambda(1)) {\bf
{\delta}}_{(\lambda;v \wedge w)}(X^{\omega_1...\omega_r}) = 0
\end {equation}
$\forall r \geq 1$. This follows by direct computation, using
Leibniz Rule (96) and the identity:
\begin {equation}
\sum_{cycl\{U,V,W\}}U.d\omega(V,W) = 0
\end {equation}
QED.}

\medskip
\medskip

Finally, let us compute the commutator $[{\bf
{\delta}}_{(\lambda;a \wedge b)},{\bf
{\delta}}_{(\eta;u \wedge v)}]$, of two tangent vectors considered as
elements of the Lie Algebra  $\widetilde {l{\cal {M}}_p}$ (see
section 3.4).

By (88), we have that:
\begin {equation}
[{\bf {\delta}}_{(\lambda;a \wedge b)},{\bf
{\delta}}_{(\eta;u \wedge v)}]={\bf
{\delta}}_{(\lambda;a \wedge b)} \star {\bf{\delta}}_{(\eta;u \wedge v)} -
{\bf{\delta}}_{(\eta;u \wedge v)} \star {\bf
{\delta}}_{(\lambda;a \wedge b)}
\end {equation}

Using (87) and (132), we can write:
\begin {eqnarray}
&&[{\bf {\delta}}_{(\lambda;a \wedge b)},{\bf
{\delta}}_{(\eta;u \wedge v)}] X^{\omega_1...\omega_r} = \epsilon \circ
\Big( \big(1 \otimes {\bf {\delta}}_{(\lambda;a \wedge b)} \big) \Delta
\big( 1 \otimes {\bf
{\delta}}_{(\eta;u \wedge v)} \big) \Delta \Big) X^{\omega_1...\omega_r}
+\nonumber \\
 && \ \ \ \ \ \ \ - \epsilon \circ  \Big( \big(1 \otimes {\bf
{\delta}}_{(\eta;u
\wedge v)} \big) \Delta  \big( 1 \otimes {\bf
{\delta}}_{(\lambda ;a \wedge b)} \big) \Delta \Big) X^{\omega_1...\omega_r}
\nonumber \\
&&\ \ \ \ \ \ \ \ \ \ \ \ \ \ = \triangle^E_{(\lambda;a \wedge b)}(\lambda(1))
\Big(  \big( 1 \otimes {\bf {\delta}}_{(\eta;u \wedge v)} \big) \Delta
X^{\omega_1...\omega_r}\Big)(\epsilon) +\nonumber \\
 && \ \ \ \ \ \ \ - \triangle^E_{(\eta;u  \wedge v)}(\eta (1)) \Big(  \big( 1
\otimes {\bf {\delta}}_{\lambda ;a \wedge b)} \big) \Delta
X^{\omega_1...\omega_r}\Big)(\epsilon)  \nonumber \\
&& \ \ \ \ \ \ \ \ \ \ \ \ \ \ = \triangle^E_{(\lambda ;a  \wedge b)}(\lambda
(1)) \Big( \sum_{i=0}^r X^{\omega_1...\omega_i}{\bf
{\delta}}_{(\eta;u \wedge v)} (X^{\omega_{i+1}...\omega_r}) \Big)(\epsilon) +
\nonumber \\
&& \ \ \ \ \ \ \ - \triangle^E_{(\eta;u  \wedge
v)}(\eta (1)) \Big( \sum_{i=0}^r X^{\omega_1...\omega_i}{\bf
{\delta}}_{(\lambda;a \wedge b)} (X^{\omega_{i+1}...\omega_r}) \Big)(\epsilon)
\nonumber \\
&&  = \sum_{i=0}^r \sum_{k=o}^i \big( {\cal D}_{a \wedge
b}(\lambda (1))X^{\omega_1...\omega_k}(\lambda)\big)\big(
X^{\omega_{k+1}...\omega_i}({\lambda}^{-1})\big) {\bf
{\delta}}_{(\eta;u \wedge v)} (X^{\omega_{i+1}...\omega_r}) + \nonumber \\
&&  - \sum_{i=0}^r \sum_{k=o}^i \big( {\cal D}_{u \wedge v}(\eta
(1))X^{\omega_1...\omega_k}(\eta)\big)\big(
X^{\omega_{k+1}...\omega_i}({\eta}^{-1})\big) {\bf
{\delta}}_{(\lambda;a \wedge b)} (X^{\omega_{i+1}...\omega_r})\nonumber \\
&&
 \end {eqnarray}
which give  the "structure constants" of $\widetilde {l{\cal {M}}_p}$.

\section {Variational Calculus}

\subsection {The action of the Diffeomorphism Group}

Denote by $Diff({\cal M})$ the Diffeomorphism group of ${\cal M}$. If
$\varphi \in Diff({\cal M})$ and $\gamma \in {\cal PM}$, we denote by
$\varphi . \gamma$ the image of the path $\gamma$ under $\varphi$.

Then we easilly prove that:

\begin {equation}
 X^{\omega_1...\omega_r}(\varphi .\gamma) =
X^{{\varphi}^{*}\omega_1...{\varphi}^{*}\omega_r}(\gamma)
\end {equation}
where $\omega_1,...,\omega_r \in
{\bigwedge}^{1}{\cal M}$ (resp., $\in C;  {\bigwedge}^{1}{\cal M} \otimes
gl(m)$).

Infinitesimally, if $\varphi_s$ is a 1-parameter group of
diffeomorphisms, with infinitesimal generator ${Y} \in {\cal XM}$,
we deduce from the previous formula, that:

\begin {eqnarray}
 D_{V}X^{\omega_1...\omega_r}({\gamma}) & \equiv &
\frac {d}{ds}{\mid}_{s=0} X^{\omega_1...\omega_r}(\varphi_s .\gamma)
\nonumber \\
&=& \sum _{i=1}^{r} X^{\omega_1...\omega_{i-1}(L_{Y}
\omega_i)\omega_{i+1}...\omega_r}(\gamma)
\end {eqnarray}
where $\omega_1,...,\omega_r \in {\bigwedge}^{1}{\cal M}$ (resp., $\in C;
{\bigwedge}^{1}{\cal M} \otimes gl(m)$), and $L_{Y}\omega$ means Lie derivative
of the 1-form $\omega$, in the direction of ${Y}$. In the above formula,
$D_{V}X^{\omega_1...\omega_r}({\gamma})$ denotes the "Fr\'echet" Derivative of
 $X^{\omega_1...\omega_r}$, at $\gamma$, in the direction of the "tangent
vector" $V=Y \circ {\gamma} \in {\gamma}^{*}T{\cal M}$, and will be treated in
detail in the next section.

 Using Cartan formula $L_{Y}=\iota_{Y}d+d\iota_{Y}$
as well as  (8-11), we can deduce the following more explicit formula:

\begin {eqnarray}
D_{V} X^{\omega_1...\omega_r}({\gamma}) &=& \sum _{i=1}^{r}
X^{\omega_1...\omega_{i-1}.({\iota_{Y}}
d\omega_i).\omega_{i+1}...\omega_r}(\gamma) \nonumber \\
 &+& \sum_{i=2}^{r} X^{\omega_1...\omega_{i-2}.\iota_{Y} (\omega_{i-1}
\wedge \omega_i).\omega_{i+1}...\omega_r}(\gamma) \nonumber \\
&+& \omega_r(V(1))X^{\omega_1...\omega_{r-1}}(\gamma) -\omega_1(V(0))
X^{\omega_2...\omega_r}(\gamma) \nonumber \\
&&
\end {eqnarray}

 \medskip
\medskip

Consider now, the "pointed" Diffeomorphism Group $Diff_p({\cal M})$, consisting
of the diffeomorphisms $\varphi$ that fix the point $p$. Its "Lie algebra"
${\cal X}_p({\cal M})$, consists of the vector fields $Y$ that vanish on $p$.

$Diff_p({\cal M})$ acts naturally on ${\cal A}_p$   through:
\begin {eqnarray}
(\varphi, X^{\omega_1...\omega_r}) &\mapsto&  \varphi.
X^{\omega_1...\omega_r}\nonumber \\
&\equiv& X^{{\varphi}^{*}\omega_1...{\varphi}^{*}\omega_r}
 \end {eqnarray}

Each such $\varphi$ is an Hopf algebra automorphism, i.e.:
\begin {equation}
\varphi.(X^{\bf u}X^{\bf v})=(\varphi.X^{\bf u}) (\varphi.X^{\bf v}) \ \ \  and
\ \ \  \Delta \circ {\varphi}=(\varphi \otimes \varphi)\circ \Delta
\end {equation}

Thus, $\varphi$ induces an automorphism of $\widetilde {{\bf L}{\cal M}_p}$,
through: \begin {eqnarray}
{\varphi}.{\tilde {\alpha}}( X^{\omega_1...\omega_r}) &\equiv& {\tilde
{\alpha}}({\varphi}. X^{\omega_1...\omega_r}) \nonumber \\
&=&{\tilde {\alpha}}(X^{{\varphi}^{*}\omega_1...{\varphi}^{*}\omega_r})
\end {eqnarray}

Considered as an element of $Aut(\widetilde {{\bf L}{\cal M}_p})$, $\varphi$
has a "differential" $d\varphi:\widetilde
{{l}{\cal M}_p} \rightarrow \widetilde {{l}{\cal M}_p}$, defined by:
\begin {equation}
d{\varphi}(\delta) (X^{\omega_1...\omega_r}) \equiv
\delta (X^{{\varphi}^{*}\omega_1...{\varphi}^{*}\omega_r})
\end {equation}
and so,  $\varphi \mapsto d\varphi$ give us a linear representation of
$Diff_p({\cal M})$ on $\widetilde {{l}{\cal M}_p}$. The
corresponding infinitesimal action  of $Y \in {\cal X}_p({\cal M})$ on
$\delta$, denoted by $Y \cdot \delta$, is given by:
\begin {equation}
\big(Y \cdot  \delta \big)(X^{\omega_1...\omega_r})
= \sum _{i=1}^{r}
{\delta}(X^{\omega_1...\omega_{i-1}.(L_{Y}\omega_i).\omega_{i+1}...\omega_r})
\end {equation}

Recall that $Y(0)=0=Y(1)$. Using this, together with Cartan formula and
relations (52-54), defining the ideal ${\bf J}_p$, we compute that:
\begin{eqnarray}
\big(Y \cdot  \delta \big)(X^{\omega_1...\omega_r})
&=& \sum _{i=1}^{r} { \delta}(X^{\omega_1...\omega_{i-1}.({\iota_{Y}}
d\omega_i).\omega_{i+1}...\omega_r}) \nonumber \\
 &+& \sum_{i=2}^{r} { \delta}(X^{\omega_1...\omega_{i-2}.\iota_{Y}
(\omega_{i-1} \wedge \omega_i).\omega_{i+1}...\omega_r})
\end {eqnarray}

\medskip

\subsection {Variational Derivative. Relation with Area Derivative. Homotopy
Invariants}

Consider a path $\gamma \in {\cal PM}_p$, based at $p$, and the "tangent space"
$T_{\gamma}{\cal PM}_p$, to ${\cal PM}_p$ at $\gamma$. By definition,
$T_{\gamma}{\cal PM}_p$ consists of sections of the pull-back bundle
$\gamma^{*}
T{\cal M}$, i.e., of vector fields along $\gamma$, that vanish on $p$. Fix a
"tangent vector" $V \in T_{\gamma}{\cal PM}_p$, and let $s \mapsto \gamma_s$ be
a curve of paths in ${\cal PM}_p$, starting at $\gamma$, in "time" $s=0$, with
velocity $V$, i.e.:

\begin {eqnarray}
\gamma_0 &=& \gamma \\
V(t) &=& \frac {\partial}{\partial s}{\mid}_{s=0} {\gamma_s}(t) \\
V(0) &=& 0.
\end {eqnarray}

We call  $s \mapsto \gamma_s$ a {\it Variation} of $\gamma=\gamma_0$,
with associated {\it Variational Vector Field} $V$. For example, in
the previous section we have consider the special case where
$\gamma_s=\varphi_s \circ \gamma$ and $V={Y} \circ \gamma$ .

 We may compute the "Fr\'{e}chet" derivative of the path functionals
$X^{\omega_1...\omega_r}$, at $\gamma \in {\cal PM}_p$. By definition,
this derivative is the linear map
$D_{.}X^{\omega_1...\omega_r}(\gamma):T_{\gamma}{\cal LM} \rightarrow
R$ (resp.,$C, gl(m)$), given by:

\begin {equation}
D_{V}X^{\omega_1...\omega_r}(\gamma) \equiv \frac {d}{d
s}{\mid}_{s=0} X^{\omega_1...\omega_r}({\gamma_s})
\end {equation}

Of course, if $V$ is induced by restriction to $\gamma$ of a vector
field ${Y} \in {\cal X}_p{\cal M}$, i.e., $V={Y} \circ \gamma$ (for
example, if $\gamma$ is embedded), then we have   the situation considered in
the previous section. In the general case, to compute the RHS of (156), we use
the following Lemma (see [H], chpt.12):

\subsubsection {Lemma}

Let $N$ be a manifold (in our case, $N$ will be $I$ or $S^1$), $\gamma:N
\rightarrow {\cal M}$ an imersion, and $\omega$ a differential form in
$\cal M$.

Assume that $\Gamma:N \times [0,\epsilon] \rightarrow {\cal M}$ is a
smooth variation of $\gamma$, with {\it Variational Vector Field} $V$.
That is, putting $\gamma_s(t)=\Gamma(t,s), \ \ \forall (t,s) \in N \times
[0,\epsilon]$, we have $\gamma_o=\gamma$ and $V(t)=\frac {\partial}{\partial
s}{\mid}_{s=0}\Gamma (t,s)=\Gamma_{{\star}_{(t,0)}}(\frac
{\partial}{\partial s}\mid_{(t,0)}), \ \ \forall t \in N$.

Then, as differential forms on $N=N \times \{0\}$ we have:
\begin {eqnarray}
\frac {d}{ds}{\mid}_{s=0} {\gamma_s}^{*}\omega &=&
{\gamma}^{*} \big({\iota}_V d{\omega} + d({\iota}_V{\omega})\big)
\nonumber \\ &=&{\gamma}^{*}({\iota}_V d{\omega}) +
d\big({\gamma}^{*}({\iota}_V{\omega})\big).
\end {eqnarray}

\medskip
\medskip
\medskip

{\underline {Notational Convention}}...{\small For each $t \in N$,
$\iota_{V(t)}\omega$ is the contraction of $\omega(\gamma(t))$ with
$V(t) \in T_{\gamma(t)}{\cal M}$. So, it's a form in
$T_{\gamma(t)}{\cal M}$. ${\gamma}^{*}({\iota}_{V(t)}{\omega})$ is
the pull-back of this form, to give a form on $T_t N$, as $t$ varies
on $N$. This defines a differential form on $N$ (denoted by
${\gamma}^{*}({\iota}_V{\omega})$, in the above formula (157)) to
which we apply $d$. This is the meaning of
$d\big({\gamma}^{*}({\iota}_V{\omega})\big)$ in formula (157). The
other term ${\gamma}^{*}({\iota}_V d{\omega})$ is similarly
interpreted, with $d\omega$ replacing $\omega$.}

\medskip
\medskip
\medskip

Now, if $\gamma:I \rightarrow {\cal M}$ is an imersed path, based at $p$,
$\gamma_s$ a variation, with variational vector field $V$, we can apply the
above Lemma to compute that:

\begin {eqnarray}
 \frac {d}{ds}{\mid}_{s=0} \Big( \int_{\gamma_s} \omega \Big) &=&
\frac {d}{ds}{\mid}_{s=0} \Big( {\int_I}{\gamma_s}^{*}\omega \Big)
\nonumber \\
    &=& \int_{I}{\gamma}^{*} \big( {\iota}_V d{\omega} +
d({\iota}_V{\omega}) \big) \nonumber \\
    &=& \int_{I} {\gamma^{*}}({\iota}_V d{\omega}) + \int_{\partial I}
{\gamma^{*}}d({\iota}_V{\omega}) \nonumber \\
     &=& \int_{I} {\gamma^{*}} ({\iota}_V d{\omega}) + \omega(V(1)) -
\omega(V(0)) \nonumber \\
     &=&\int_{\gamma} {\iota}_V d{\omega} + \omega(V(1))
\end {eqnarray}
where in the last equality we have used the  notation $\int_{\gamma} {\iota}_V
d{\omega}$ for $\int_{I} {\gamma^{*}} ({\iota}_V d{\omega})$.

In particular, for a loop $\gamma \in {\cal LM}_p$, using that notation and the
fact that $V(0)=0=V(1)$ we have:

\begin {equation}
D_{V}X^{\omega}(\gamma) = X^{{\iota}_V
d{\omega}}(\gamma) =\int_{\gamma}{\iota}_V d{\omega}.
\end {equation}

Finally, using Lemma 5.2.1, (8-9) and induction, we can deduce that:

\begin {eqnarray}
D_{V} X^{\omega_1...\omega_r}(\gamma) &=& \sum _{i=1}^{r}
\int_{\gamma}\omega_1...\omega_{i-1}.{\iota_{V}}
(d\omega_i).\omega_{i+1}...\omega_r \nonumber \\
 &+& \sum_{i=2}^{r} \int_{\gamma}\omega_1...\omega_{i-2}.\iota_{V}
(\omega_{i-1} \wedge \omega_i).\omega_{i+1}...\omega_r \nonumber \\
&+&  \Big(\int_{\gamma}\omega_1...\omega_{r-1}\Big).\omega_r(V(1))
\end {eqnarray}
where we used the above mentioned notational conventions.

For an imersed loop $\gamma \in {\cal LM}_p$, we consider the restricted class
of variations $V$, that keep the base point $p$ fixed:
\begin {equation}
{\cal V}_p \equiv \{V \in {\gamma}^{*}T{\cal M} : \ \ V(0)=0=V(1)\}
\end {equation}

Any solution $\Psi$ of the equation:
\begin {equation}
D_{\gamma} \Psi (V) =0 \ \ \ \ \ \ \ \  \forall V \in {\cal V}_p
  \end {equation}
is called a (relative) {\it Homotopy Invariant} of the loop $\gamma$.

\medskip

  It's useful, in this moment, to analyze the relationship between the above
mentioned "Fr\'echet" Derivative and the Area Derivative introduced in section
4.2. Thus, define an element of $\widetilde {{l}{\cal M}_p}$, by:

\begin {equation}
\Theta(\gamma;V) \equiv  {\int}_0^1 \delta_{({\gamma}_o^t;V(t) \wedge
\dot{\gamma}(t))}(\gamma(t)) dt
\end {equation}
 where $V \in {\cal V}_p$ and ${\gamma}_o^t$ denotes the portion of
$\gamma$, from ${\gamma}(0)$ to ${\gamma}(t)$. The meaning of the above
expression is, as usual, the following: for each $X^{\bf u}$: \begin {equation}
\Theta(\gamma;V) (X^{\bf u}) \equiv  {\int}_0^1 \delta_{({\gamma}_o^t;V(t)
\wedge \dot{\gamma}(t))}(\gamma(t)) (X^{\bf u}) dt
\end {equation}
if, of course, the RHS is well defined.

Now,  using the notations of section 4.2, we can prove, after a tedious
computation, the following formula (see figure 7):
\begin {eqnarray}
 D_{V} X^{\bf u} (\gamma) &=& \int_0^1 \triangle_{({\gamma}_o^t;V \wedge
\dot{\gamma})}(\gamma(t)) X^{\bf u} dt \\
&=& \gamma \circ \Big( \Theta(\gamma;V) \otimes 1 \Big) \circ \Delta X^{\bf
u}
\end {eqnarray}

\begin{figure}[htb]
\vspace{1in}
\caption{ $ \int_{\gamma} \Delta_{({\gamma}_o^t;V \wedge
\dot{\gamma})}(\gamma(t))dt $}.
\end{figure}

For example, if $U:{\bf L}{\cal M}_p \rightarrow gl(n;R)$ is the holonomy of a
connection $\omega$ on the trivial bundle $R^n \times {\cal M} \rightarrow
{\cal M}$ (see section 2), then, using  the above formulas, we can compute
that:
\begin {equation}
D_{ V} U (\gamma) = U_{\gamma}. \Big(\int_0^1   U_{{\gamma}_0^t}
\Omega_{\gamma (t)} \big( V(t) \wedge \dot{\gamma}(t)\big).
 U_{({\gamma}_0^t)^{-1}} \Big)
\end {equation}
a formula that it's called sometimes the "Non-Abelian Stokes Theorem" (see
[FGK]).

\medskip

Incidently, the same computations show that:

\begin {equation}
{\bf{\delta}}_{(\lambda;u \wedge v)} U_{\lambda} =  U_{\lambda}
{\Omega}_{\lambda(1)}(u \wedge v).  U_{{\lambda}^{-1}}
\end {equation}

\medskip
 \medskip
\medskip

Using  definition (162) as well as the above
"Non-abelian Stokes Theorem" (167), we can produce, in a constructive way,
several such homotopy invariants, which can be useful for example for loop
representation of Chern-Simmons Theory. Let us detail this point.

 \medskip
\medskip

{\small Consider  a system of 1-forms on $\cal M$, of type:

$$\begin{array}{cccccc}
\omega_1 & \omega_2 & \omega_3 & \ldots &\ldots & \omega_r \\
\omega_{12}  &  \omega_{23}  & \omega_{34}        &  \ldots & \ldots &
\omega_{r-1 \,r}         \\
 \omega_{123}   &     \omega_{234}      & \omega_{345} & \ldots      &
\omega_{r-2 \, r-1\,r}  &    \\
 \ldots  & \ldots & \ldots  &   &    &     \\
\omega_{12...r}   &           &
  &        &     &     \end{array}$$
and, with them, construct    the nilpotent connection 1-form (see section 2):
$$\omega = \left[
\begin{array}{cccccc}
0  &  \omega_1  & \omega_{12}        &  \omega_{123} & \ldots &
\omega_{12...r}       \\
 0  &     0      & \omega_2 & \omega_{23}      & \ldots & \omega_{2...r}
\\
 \ldots  & \ldots & \ldots  & \ldots & \ldots  & \ldots   \\
0  &     0      &    0     &   \ldots      &    0    &  \omega_r \\
0  &    0      &    0     &   \ldots      &    0    & 0
\end{array}
\right]$$

The holonomy of this connection is given by:
$$U = \left[
\begin{array}{cccccc}
1   &  \int {\bf u}_1  & \int {\bf u}_{12} & \int  {\bf u}_{123}& \ldots & \int
 {\bf u}_{12...r} \\
 0  &     1      & \int {\bf u}_2 & \int {\bf u}_{23}  & \ldots & \int
 {\bf u}_{23...r} \\
  0 &       0      & 1   &  \int {\bf u}_3 &  \ldots   & \int  {\bf
u}_{3...r}\\ \ldots & \ldots &\ldots &\ldots &\ldots &\ldots \\
 0 &     0       &    0     & \ldots  & 1  & \int {\bf u}_r \\
0 & 0 & 0 & \ldots  & 0  & 1
\end{array}
\right]$$
where:
\begin {eqnarray*}
{\bf u}_1&=&\omega_1 \nonumber\\
{\bf u}_{12}&=&\omega_1  \omega_2 + \omega_{12} \nonumber\\
{\bf u}_{123}&=& \omega_1  \omega_2 \omega_3+ \omega_{12}\omega_3 +
\omega_{1}\omega_{23} + \omega_{123} \nonumber\\
 & ........& \nonumber \\
{\bf u}_{12...r} &=& \omega_1...\omega_r + \omega_{12}\omega_{3...r} +... +
\omega_{12...r}
\nonumber \\
{\bf u}_2&=&\omega_2 \nonumber\\
{\bf u}_{23}&=&\omega_2  \omega_3 + \omega_{23} \nonumber\\
& ........& \nonumber \\
\end {eqnarray*}
and the curvature of $\omega$ is:
$$\Omega = \left[
\begin{array}{cccccc}
0  &  W_1  & W_{12}        &  W_{123} & \ldots &
W_{12...r}       \\
 0  &     0      & W_2 & W_{23}      & \ldots & W_{2...r}
\\
 \ldots  & \ldots & \ldots  & \ldots & \ldots  & \ldots   \\
0  &     0      &    0     &   \ldots      &    0    &  W_r \\
0  &    0      &    0     &   \ldots      &    0    & 0
\end{array}
\right]$$
where:
\begin{eqnarray}
W_1&=&d\omega_1   \nonumber \\
W_{12}&= & \omega_{1} \wedge \omega_{2}+d\omega_{12} \nonumber \\
 W_{123} &= &  \omega_{1} \wedge \omega_{23}+\omega_{12} \wedge
\omega_{3}+d\omega_{123} \nonumber  \\
 & & ................................ \nonumber  \\
W_{12...r} &=& \omega_{1} \wedge \omega_{2...r}+\omega_{12} \wedge
\omega_{3...r}+...+d\omega_{12...r} \nonumber \\
W_2&=&d\omega_2 \nonumber \\
W_{23}&= & \omega_{2} \wedge \omega_{3}+d\omega_{23} \nonumber \\
& & ................................ \nonumber
\end{eqnarray}}

As is well known (see [KN]) the holonomy $U$ is a homotopy invariant, iff the
curvature is zero. In fact, this can be proved directly, using equation (162)
as well as the  nonabelian Stokes theorem (167). So, forcing the curvature
$\Omega$ to be $0$, i.e., forcing the above 2-forms $W_{12...r}$ to be $0$, we
have that each entry in the above matrix $U$ gives an homotopy invariant.

\medskip

  To construct such a {\it Flat Nilpotent Connection}, assume that
$H^2({\cal M},R)=0$ and start with closed 1-forms $\omega_1,...,\omega_r$.
Then,  $\omega_1 \wedge \omega_2$,...,$\omega_{r-1} \wedge \omega_r$ are also
closed 2-forms and we can solve the  equations obtained equalizing to zero each
entry of the matrix $\Omega$, for
$\omega_{12}$,...,$\omega_{r-1\,r}$,...,$\omega_{1...r}$, step by step.

In fact, since $\omega_1 \wedge \omega_2$,...,$\omega_{r-1} \wedge \omega_r$
are
closed 2-forms, they are also exact, and thus, we can find
$\omega_{12},...,\omega_{r-1\,r}$. Now:
\begin {eqnarray*}
d(\omega_{1} \wedge \omega_{23}+\omega_{12} \wedge
\omega_{3})=\omega_{1} \wedge \omega_{2} \wedge \omega_{3}-\omega_{1} \wedge
\omega_{2} \wedge \omega_3 =0
\end {eqnarray*}
and  we can find $\omega_{123}$, and so on.

\medskip
\medskip

 {\bf Acknowledgments}

I am extremelly grateful to J.Mour\~ao for many helpful comments and criticisms
which substantially improved the presentation. I also thank the useful
comments of A. Ashtekar, A. Caetano, R. Gambini, R. Loll,
Yu.M. Makeenko, C.M. Menezes, R.
Picken,  E. Rego and  C. Rovelli.

\medskip
\medskip

{\Large \bf References}

\medskip

[Ab]. Abe E.. {\it Hopf Algebras}, (1977) Cambridge Univ. Press

\medskip

[Ash]. Ashtekar A.. {\it Lectures on Nonperturbative Canonical
Gravity}, (1992) World Scientific.

\medskip

[B]. Barrett J.W.. Holonomy Description of Classical Yang-Mills Theory and
General Relativity. {\it Int.J. of Theor. Phys.} 30 (1991) 1171.

\medskip

[BGG]. Di Bartolo C., Gambini R., Griego J.. The Extended Loop Group: an
Infinite Dimensional Manifold associated with the Loop Space. {\it Preprint }
(1992).

\medskip

[Chen1]. Chen K.T.. Iterated Path Integrals. {\it Bull. A.M.S.} 83
(1977) 831-881.

\medskip

[Chen2]. Chen K.T.. Iterated Integrals of Differential Forms and Loop
Space Homology. {\it Ann. Math.} 97 (1973) 217-243.

\medskip

[Chen3]. Chen K.T.. Integration of Paths: a Faithfull Representation
of Paths by Noncommutative Formal Power Series. {\it Trans. A.M.S.}
89 (1958) 385-407.

\medskip

[Chen4]. Chen K.T.. Algebras of Iterated Path Integrals and
Fundamental Groups. {\it Trans. A.M.S.} 156 (1971) 359-379.

\medskip

[Chen5]. Chen K.T.. Algebraic Paths. {\it Journal of Algebra} 9 (1968), 8-36.

\medskip

[DF]. Dollard J.D., Friedman C.N.. {\it Product Integration with Applications
to Differential Equations} Reading; Addison Wesley 1979.

\medskip

[FGK]. Fishbane P.M., Gasiorowicz S., Kaus P.. Stokes Theorem for Nonabelian
Fields {\it Phys. Review D} 24 (1981) 2324-2329.

\medskip

[GT1]. Gambini R., Trias A.. Geometric Origin of Gauge Theories. {\it Phys.
Review D} 23 (1981) 553-555.

\medskip

[GT2]. Gambini R., Trias A.. Gauge Dynamics in the C Representation. {\it
Nuclear Physics B} 278 (1986) 436-448.

\medskip

[GT3]. Gambini R., Trias A.. Path Dependent Formulation of Gauge Theories and
the Origin of the Field Copies in the Non-Abelian Case. {\it Phys. Review D}
21, 12 (1980) 3413-3416.

\medskip

[G1]. Gambini R.. {\it Teorias de Calibre en el Espacio de Ciclos}. Univ. Simon
Bolivar. Mimeog. Notes.

\medskip

[G2]. Gambini R.. Loop Space Representation of Quantum General Relativity and
the Group of Loops. {\it Phys. Letters B} 255 (1991) 180-188.

\medskip

[H]. Hermann R.. {\it Differential Geometry and the Calculus of Variations}.
Academic Press, 1968.

\medskip

[KN]. Kobayashi S., Nomizu K.. {\it Foundations of Differential Geometry. Vol
1} Interscience Publ. John Wiley, 1963.

\medskip

[Lew]. Lewandowsky J.. Group of Loops, Holonomy Maps and Path Bundle. {\it
Preprint} 1992.

\medskip

[L]. Loll R.. Loop Approaches to Gauge Field Theory. {\it Syracuse Preprint}
1992.

\medskip

[Man]. Mandelstam . Quantum Electrodynamics without Potentials. {\it Ann.
Phys. (NY)} 19 (1962) 1-24.

\medskip

[MM]. Makeenko Yu.M., Migdal A.A.. Quantum Chromodynamics as Dynamics of
Loops. {\it Nuclear Phys. B} 188 (1981)882-893.

\medskip

 [Mal]. Mallios A.. {\it Topological Algebras.
Selected Topics}. North-Holland Mathematics Studies 124. North-Holland.

\medskip

[RS]. Rovelli C., Smolin L.. Loop Space Representation of Quantum General
Relativity. {Nuclear Phys. B} 331 (1990) 80-152.

\medskip

[R]. Rovelli c.. Ashtekar Formulation of General Relativity and the Loop Space
Non-Perturbative Quantum Gravity: A Repport. {\it Class. Quantum Grav.} (1991)
1613-1675.

\medskip

[Sw]. Sweedler M.E.. {\it Hopf Algebras}, 1969. W.A. Benjamin, Inc. Publishers.

\medskip

[T]. Teleman M.C.. G\'en\'eralization du Groupe Fondamental. {\it Ann. Scient.
\'Ec. Norm. Sup.} t.77 (1960) 195-234.

\end{document}